\documentclass[longauth]{aa} 

\usepackage{graphicx,subcaption}
\usepackage{txfonts}
\usepackage{natbib}

\begin{document}

   \title{The broad-lined Type-Ic supernova SN 2022xxf with extraordinary two-humped light curves}

   \subtitle{I. Signatures of H/He-free interaction in the first four months}

   \author{
H.~Kuncarayakti\inst{1,2},
J.~Sollerman\inst{3},
L.~Izzo\inst{4},
K.~Maeda\inst{5},
S.~Yang\inst{3,6},
S.~Schulze\inst{7},
C.~R.~Angus\inst{4},
M.~Aubert\inst{8},
K.~Auchettl\inst{9,10,11},
M.~Della~Valle\inst{12},
L.~Dessart\inst{13},
K.~Hinds\inst{14},
E.~Kankare\inst{1,15},
M.~Kawabata\inst{5,49},
P.~Lundqvist\inst{3},
T.~Nakaoka\inst{16},
D.~Perley\inst{14},
S.~I.~Raimundo\inst{4,17,18},
N.~L.~Strotjohann\inst{19},
K.~Taguchi\inst{5},
Y.-Z.~Cai\inst{20,21,50},
P.~Charalampopoulos\inst{1},
Q.~Fang\inst{5},
M.~Fraser\inst{22},
C.~P.~Guti\'errez\inst{2,1},
R.~Imazawa\inst{23},
T.~Kangas\inst{2,1},
K.~S.~Kawabata\inst{16},
R.~Kotak\inst{1},
T.~Kravtsov\inst{24,1},
K.~Matilainen\inst{25,1},
S.~Mattila\inst{1,26},
S.~Moran\inst{1},
I.~Murata\inst{5},
I.~Salmaso\inst{27,28},
J.~P.~Anderson\inst{24,29},
C.~Ashall\inst{30},
E.~C.~Bellm\inst{31},
S.~Benetti\inst{27},
K.~C.~Chambers\inst{32},
T.-W.~Chen\inst{33,34},
M.~Coughlin\inst{35},
F.~De~Colle\inst{36},
C.~Fremling\inst{37,38},
L.~Galbany\inst{39,40},
A.~Gal-Yam\inst{41},
M.~Gromadzki\inst{42},
S.~L.~Groom\inst{43},
A.~Hajela\inst{4},
C.~Inserra\inst{44},
M.~M.~Kasliwal\inst{37},
A.~A.~Mahabal\inst{37,45},
A.~Martin-Carrillo\inst{22},
T.~Moore\inst{46},
T.~E.~M\"uller-Bravo\inst{39,40},
M.~Nicholl\inst{46},
F.~Ragosta\inst{47,51},
R.~L.~Riddle\inst{38},
Y.~Sharma\inst{37},
S.~Srivastav\inst{46},
M.~D.~Stritzinger\inst{48},
A.~Wold\inst{43},
D.~R.~Young\inst{46}
          }
             
   \institute{
Tuorla Observatory, Department of Physics and Astronomy, FI-20014 University of Turku, Finland
\and Finnish Centre for Astronomy with ESO (FINCA), FI-20014 University of Turku, Finland 
\and Department of Astronomy, Oskar Klein Centre, AlbaNova, 10691 Stockholm, Sweden    
\and DARK, Niels Bohr Institute, University of Copenhagen, Copenhagen, Denmark
\and Department of Astronomy, Kyoto University, Kitashirakawa-Oiwake-cho, Sakyo-ku, Kyoto, 606-8502, Japan 
\and Henan Academy of Sciences, Zhengzhou 450046, Henan, China
\and The Oskar Klein Centre, Department of Physics, Stockholm University, AlbaNova University Center, SE 106 91 Stockholm, Sweden
\and Universit\'e Clermont Auvergne, CNRS/IN2P3, LPC, Clermont-Ferrand, France
\and School of Physics, The University of Melbourne, Parkville, VIC 3010, Australia
\and ARC Centre of Excellence for All Sky Astrophysics in 3 Dimensions (ASTRO 3D), Sydney 2006, Australia
\and Department of Astronomy and Astrophysics, University of California, Santa Cruz, CA 95064, USA
\and Capodimonte Astronomical Observatory, INAF-Napoli, Salita Moiariello 16, I-80131 Napoli, Italy
\and Institut d'Astrophysique de Paris, CNRS-Sorbonne Universit\'e, 98 bis boulevard Arago, F-75014 Paris, France
\and Astrophysics Research Institute, Liverpool John Moores University, IC2, Liverpool Science Park, 146 Brownlow Hill, Liverpool L3 5RF, UK
\and Turku Collegium for Science, Medicine and Technology, University of Turku, FI-20014 Turku, Finland
\and Hiroshima Astrophysical Science Center, Hiroshima University, 1-3-1 Kagamiyama, Higashi-Hiroshima, Hiroshima 739-8526, Japan
\and Department of Physics and Astronomy, University of California, Los Angeles, CA, USA
\and Department of Physics \& Astronomy, University of Southampton, Southampton, UK
\and Benoziyo Center for Astrophysics, The Weizmann Institute of Science, Rehovot, 76100, Israel
\and Yunnan Observatories, Chinese Academy of Sciences, Kunming 650216, PR China
\and Key Laboratory for the Structure and Evolution of Celestial Objects, Chinese Academy of Sciences, Kunming 650216, PR China
\and School of Physics, O'Brien Centre for Science North, University College Dublin, Belfield, Dublin 4, Dublin, Ireland
\and Department of Physics, Graduate School of Advanced Science and Engineering, Hiroshima University, Kagamiyama, 1-3-1 Higashi-Hiroshima, Hiroshima 739-8526, Japan
\and European Southern Observatory, Alonso de C\'ordova 3107, Casilla 19, Santiago, Chile
\and Nordic Optical Telescope, Aarhus Universitet, Rambla Jos\'e Ana Fern\'andez P\'erez 7, local 5, E-38711 San Antonio, Bre\~na Baja Santa Cruz de Tenerife, Spain
\and School of Sciences, European University Cyprus, Diogenes street, Engomi, 1516 Nicosia, Cyprus
\and INAF - Osservatorio Astronomico di Padova, Vicolo dell'Osservatorio 5, 35122, Padova, Italy
\and Universit\`a degli Studi di Padova, Dipartimento di Fisica e Astronomia, Vicolo dell'Osservatorio 2, 35122, Padova, Italy
\and Millennium Institute of Astrophysics MAS, Nuncio Monsenor Sotero Sanz 100, Off. 104, Providencia, Santiago, Chile
\and Department of Physics, Virginia Tech, Blacksburg, VA 24061, USA
\and DIRAC Institute, Department of Astronomy, University of Washington, 3910 15th Avenue NE, Seattle, WA 98195, USA
\and Institute for Astronomy, University of Hawaii
\and Technische Universit{\"a}t M{\"u}nchen, TUM School of Natural Sciences, Physik-Department, James-Franck-Stra{\ss}e 1, 85748 Garching, Germany
\and Max-Planck-Institut f{\"u}r Astrophysik, Karl-Schwarzschild Stra{\ss}e 1, 85748 Garching, Germany
\and School of Physics and Astronomy, University of Minnesota, Minneapolis, Minnesota 55455, USA
\and Instituto de Ciencias Nucleares, Universidad Nacional Aut\'onoma de M\'exico, A. P. 70-543, 04510 D.F. Mexico, Mexico
\and Division of Physics, Mathematics, and Astronomy, California Institute of Technology, Pasadena, CA 91125, USA
\and The Caltech Optical Observatories, California Institute of Technology, Pasadena, CA 91125, USA
\and Institute of Space Sciences (ICE, CSIC), Campus UAB, Carrer de Can Magrans, s/n, E-08193 Barcelona, Spain.
\and Institut d'Estudis Espacials de Catalunya (IEEC), E-08034 Barcelona, Spain.
\and Department of Particle Physics and Astrophysics, Weizmann Institute of Science, Rehovot, Israel
\and Astronomical Observatory, University of Warsaw, Al. Ujazdowskie 4, 00-478 Warszawa, Poland
\and IPAC, California Institute of Technology, 1200 E. California Blvd, Pasadena, CA 91125, USA
\and Cardiff Hub for Astrophysics Research and Technology, School of Physics \& Astronomy, Cardiff University, Queens Buildings, The Parade, Cardiff, CF24 3AA, UK
\and Center for Data Driven Discovery, California Institute of Technology, Pasadena, CA 91125, USA
\and Astrophysics Research Centre, School of Mathematics and Physics, Queen’s University Belfast, Belfast BT7 1NN, UK 
\and INAF - Osservatorio Astronomico di Roma, Via Frascati 33 I-00078 Monte Porzio Catone (Roma), Italy
\and Department of Physics and Astronomy, Aarhus University, Ny Munkegade 120, DK-8000 Aarhus C, Denmark
\and Nishi-Harima Astronomical Observatory, Center for Astronomy, University of Hyogo, 407-2 Nishigaichi, Sayo-cho, Sayo, Hyogo 679-5313, Japan
\and  International Centre of Supernovae, Yunnan Key Laboratory, Kunming 650216, P.R. China
\and Space Science Data Center - ASI Via del Politecnico snc, I-00133 Roma, Italy
             \\
             }


   \date{Accepted version, 7 Aug 2023}

 
  \abstract
   {We report on our study of supernova (SN) 2022xxf based on observations obtained during the first four months of its evolution. The light curves (LCs) display two humps {of} similar maximum brightness separated by 75 days, unprecedented for a broad-lined (BL) Type Ic supernova (SN IcBL). SN~2022xxf is the most nearby SN IcBL to date (in NGC~3705, $z = 0.0037$, {at a distance of about} 20 Mpc).
   Optical and near-infrared photometry and spectroscopy are used to identify the energy source powering the LC. Nearly 50 epochs of high signal-to-noise-ratio spectroscopy were obtained within 130 days, comprising an unparalleled dataset for a SN IcBL, and one of the best-sampled SN datasets to date.
   The global spectral appearance and evolution of SN~2022xxf points to typical SN Ic/IcBL, with broad features (up to $\sim14000$ km~s$^{-1}$) and a gradual transition from the photospheric to the nebular phase. However, narrow emission lines (corresponding to $\sim1000-2500$ km~s$^{-1}$) are present {in the spectra} from the time of the second rise, suggesting slower-moving circumstellar material (CSM). These lines are subtle, in comparison to the typical strong narrow lines of CSM-interacting SNe, for example, Type IIn, Ibn, and Icn, but some are readily noticeable at late times such as in Mg~I~$\lambda$5170 and [O~I]~$\lambda$5577.
   Unusually, the near-infrared spectra show narrow line peaks {in a number of} features formed by ions of O and Mg.
   We infer the presence of CSM that is free of H and He. We propose that the radiative energy from the ejecta-CSM interaction is a plausible explanation for the second LC hump. This interaction scenario is supported by the color evolution, which progresses to the blue as the light curve evolves along the second hump, and the slow second rise and subsequent rapid LC drop.
   SN~2022xxf may be related to an emerging number of CSM-interacting SNe Ic, which show slow, peculiar LCs, blue colors, and subtle CSM interaction lines. The progenitor stars of these SNe likely experienced an episode of mass loss shortly prior to explosion consisting of H/He-free material.
   }


\titlerunning{SN 2022xxf}
\authorrunning{Kuncarayakti et al.}

   \maketitle
%

\section{Introduction}

The demise of massive stars ($M_{\textrm{ZAMS}}\gtrsim8-10~M_\odot$) as core-collapse (CC) supernovae (SNe) comes in various flavors \citep[see e.g.][]{langer12,galyam17}. {SN diversity is thought to be mainly affected by initial mass and mass loss experienced by the progenitor star.} 
Hydrogen-poor, stripped envelope (SE) SNe originate from progenitors that have lost a significant part of their envelopes before the explosion. 
These include SNe Type IIb (He-rich, little H), Ib (He-rich, no H), and Ic (no H nor He).
Significant mass loss, e.g. through strong stellar winds or interaction with a close binary companion, is required for a star to become a SESN progenitor. Such strong winds are expected for very massive progenitors ($\gtrsim30~M_\odot$, e.g. ~\citealp{crowther07,Groh2013}), while in the binary scenario the progenitors can be of relatively lower mass ($\lesssim20$~$M_\odot$,~\citealp[e.g.,][]{Yoon2015,dessart20}). 
Evidence is mounting from both studies of individual objects and samples (e.g. \citealp{taddia15,lyman16,kangas17,fang2019,prentice19}) that binaries play an important role in producing SESN progenitors.

In common SESNe, evidence for the presence of significant circumstellar material (CSM) from progenitor mass loss is rare, but has been found in some objects, e.g. late-time broad flat-topped H$\alpha$ emission in a few Type IIb SNe \citep{Matheson2000,maeda2015,fremling19}. 
In the radio and X-ray wavelengths, signatures of CSM are more frequently detected \cite[e.g.][]{horesh20}.
SNe 2014C \citep{milisav15}, 2017ens \citep{chen18}, 2017dio \citep{hk18a}, and 2018ijp \citep{tartaglia21} constitute cases where Type Ib/c SESNe spectroscopically metamorphosed into CSM-interacting Type IIn supernovae, revealing the presence of H-rich external CSM.
A small number of Type Ic SNe have been shown to interact with H/He-poor CSM, such as SNe 2010mb \citep{benami14} and 2021ocs \citep{hk22}. They show slow light curves with blue colors and distinct emission lines due to the CSM interaction.
\cite{sollerman20} presented two SESNe (2019oys and 2019tsf) {which starts to re-brighten after a few months of the light curve peaks}. 
They concluded that the extra power needed for such a light curve (LC) evolution is presumably CSM interaction because none of the other powering mechanisms at play at later phases are likely to result in such a behavior. However, only SN 2019oys showed clear evidence for such interaction (e.g. narrow coronal lines) whereas SN 2019tsf did not.
{To account for the bumpy LCs seen in SN 2019tsf, a scenario involving interaction with} a warped CSM disk influenced by a tertiary companion was suggested \citep{zenati22}.
Interaction between a newborn neutron star and a binary companion star has also been proposed to produce bumpy SN LCs \citep{hirai22}.
LC bumps are relatively common in {Type I luminous and superluminous SNe}. They have been suggested to be caused by central engine (e.g. magnetar) activities, CSM interaction, or combination of both {\citep[e.g.][]{gomez21,hosseinzadeh22,moriya22,chen23,lin23}}.

In this paper {we present the observations of SN 2022xxf, an SESN with a spectacular second hump in its light curves.}\footnote{In the literature, the term ``double-peaked LC'' is used indiscriminately for objects with early, fast-declining shock cooling emission e.g. SN~1993J \citep{richmond94}, or other types of slower peaks due to other mechanisms e.g. SN~2005bf \citep{folatelli06}. Here we chose the word `hump' for SN~2022xxf as the rise and fall phases are well observed, forming roundish shapes, and the ease of association with the shape of a bactrian camel's back with the characteristic two humps.}
SN 2022xxf was discovered in NGC~3705 by \citet{itagaki22} on 2022-10-17 ($\mathrm{MJD_{\rm{discovery}}=59869.85}$) in a white-light image, and was reported to the Transient Name Server (TNS) on the same day. The host has a redshift of $z = 0.00340 \pm 0.00001$, distance modulus $\mu = 31.54 \pm 0.45$ mag, and luminosity distance $20.3^{+4.7}_{-3.8}$ Mpc \citep[][via NASA/IPAC Extragalactic Database\footnote{\url{http://ned.ipac.caltech.edu/}}]{tully16}. From measurements of host galaxy lines at the line of sight towards the SN (Sect.~\ref{sec:spectro}), we estimate and adopt a redshift of $z = 0.0037$ for the SN, which is used to correct the spectra.
Spectral classification as a Type-IcBL SN was reported by \citet{balcon22}, and confirmed with a spectrum taken earlier by our group (\citealt{nakaoka22}; see Sect. \ref{sec:spectro}).
\citet{corsi22} reported a radio detection at 5.5 GHz using the Very Large Array (VLA).

\section{Observations and data reduction}
\label{sec:obs}

Observations of SN~2022xxf were done with a number of facilities, as listed in Table~\ref{tab:instru} which includes the instrument references.

\subsection{Photometry}

The first observations of SN~2022xxf/ZTF22abnvurz with the Zwicky Transient Facility \cite[ZTF;][]{Graham2019,Bellm2019}, and the Palomar Schmidt 48-inch (P48) Samuel Oschin telescope under the twilight survey \citep{Bellm2019scheduler}, were made on 2022-10-18 ($\mathrm{MJD^{\rm{ZTF}}_{\rm{first\ detection}}=59870.53}$) in \textit{r}-band, one day after the discovery. Photometry was obtained via the ZTF forced photometry service\footnote{\url{https://ztfweb.ipac.caltech.edu/cgi-bin/requestForcedPhotometry.cgi}} \citep{masci19}.
No immediate pre-explosion non-detections are available as the SN just emerged from solar conjunction.
After peak, additional \textit{g} and \textit{i}-band data were obtained with the ZTF camera on the P48 and Spectral Energy Distribution Machine (SEDM) Rainbow Camera on the Palomar 60-inch telescope.
The P60 data were reduced using FPipe \citep{Fremling2016FPipe} for image subtraction. 
Further, we complemented the above with photometry from the Liverpool Telescope (LT), IO:O camera with \textit{ugriz} filters. An automatic pipeline reduces the images, performing bias subtraction, trimming of the overscan regions, and flat fielding. Template subtraction was done for the photometry.
Photometry was also performed for the images taken with the 3.8-m Seimei telescope \citep{Kurita2020} at the Okayama Observatory, Kyoto University, using the TriColor CMOS Camera and Spectrograph (TriCCS). 
Near-infrared (NIR) photometry was obtained using NOTCam at the 2.56-m Nordic Optical Telescope (NOT) at the Observatorio del Roque de los Muchachos on La Palma (Spain), and SOFI at the ESO New Technology Telescope (NTT) in La Silla, Chile.
Standard reduction with bias subtraction and flat fielding was performed. No image subtraction was performed for the TriCCS and NIR data due to the lack of reference images.

\subsection{Spectroscopy}
\label{sec:spectro}

The first spectrum of SN~2022xxf was obtained using the Hiroshima One-shot Wide-field Polarimeter (HOWPol) on the 1.5-m Kanata telescope at the Higashi-Hiroshima Observatory, Hiroshima University. 
{Based on this spectrum, this SN was classified as a broad-lined SN Ic} \citep{nakaoka22}.
Within ZTF, a series of spectra were obtained with the SEDM and reduced with the pipeline described by \citet{Rigault2019}, {while some were collected} with the Double Beam Spectrograph (DBSP) on the Palomar 200-in telescope, reduced using a DBSP reduction pipeline \citep{mandigo22} relying on PypeIt \citep{Prochaska2020}.
Spectra were also obtained with the Alhambra Faint Object Spectrograph and Camera (ALFOSC) using grism \#4 on the NOT, by the ZTF and NUTS\footnote{\url{http://nuts2.sn.ie/}} collaborations. ALFOSC data reductions were performed using ALFOSCGUI\footnote{\url{https://sngroup.oapd.inaf.it/foscgui.html}}.
We obtained optical spectra using EFOSC2, and near-infrared spectra using SOFI, both at ESO NTT, as part of the ePESSTO+ survey \citep{Smartt2015}. The raw data were reduced using the dedicated PESSTO data reduction pipeline\footnote{\url{https://github.com/svalenti/pessto}}.
Kyoto Okayama Optical Low-dispersion Spectrograph with optical-fiber Integral Field Unit (KOOLS-IFU) was also used for spectroscopy, using the VPH-blue grism.
The data reduction was performed with the Hydra package in IRAF and dedicated software\footnote{\url{http://www.o.kwasan.kyoto-u.ac.jp/inst/p-kools/reduction-201806/index.html}}.
All the spectroscopic observations were accompanied with standard star observations and followed by standard reductions which include bias and flat corrections, and wavelength and flux calibrations.

In addition to {the above} low-resolution {spectra}, we also obtained {an} intermediate-resolution {spectrum with the X-Shooter mounted on the} ESO Very Large Telescope (VLT) in Cerro Paranal, Chile, on 2022-12-18\footnote{ESO Director's Discretionary Time, program 110.25A0.001, PI: Izzo.}. The observations were performed using the standard nod-on-slit mode, but each single arm spectrum was reduced using the ``stare'' mode reduction, given the  brightness of  the SN, and finally stacked using the standard X-Shooter pipeline \citep{Goldoni2006,Modigliani2010}. Residual sky lines have been interpolated using the background as reference, and finally a telluric correction was implemented using  the line-by-line radiative transfer model (LBLRTM;
\citealp{Clough1992}).

\section{Results and discussion}
\label{sec:discussion}

\subsection{Extinction, metallicity, and redshift}
The amount of reddening in SN~2022xxf is estimated using the host galaxy narrow Na~I~D absorption lines in the intermediate-resolution X-Shooter spectrum. This yields Na~I (D1+D2) equivalent width of $\sim 1.5$~\AA~ from the line measurements through Voigt profile fitting. Employing the relations from \citet{poznanski12}, this equivalent width corresponds to {a reddening} of $E(B-V) = 0.8 \pm 0.2$ mag, assuming $R_V = 3.1$. 
We note, however, that in the range of the measured equivalent width the \citeauthor{poznanski12} relation is not well sampled. Furthermore, any presence of CSM could have contributed to the Na~I absorption and it could have a different $R_V$, complicating the extinction estimate.
This caveat should be kept in mind where extinction is a factor in the analysis.
The host galaxy of SN~2022xxf is inclined\footnote{Estimated using \url{https://edd.ifa.hawaii.edu/inclinet/}} at about 68$\degr$, which makes high extinction likely in any case. 
The foreground Milky Way extinction is negligible, $E(B-V)_{MW} = 0.04$ mag \citep{schlafly11}, which is confirmed by our measurements of the Na~I~D lines from the Milky Way in the X-Shooter spectrum. We thus assume $E(B-V) = 0.8 \pm 0.2$ mag as the total line of sight extinction for SN~2022xxf throughout the paper, unless mentioned otherwise.

From detected interstellar medium lines of H$\alpha$ and [N~II] $\lambda$6584, the strong line method yields near-solar metallicity 12+log(O/H) = $8.56\pm0.16$ dex (N2, \citealt{marino13}), although the strong presence of the SN spectrum prevents the determination of the host stellar population properties with certainty.
This metallicity is considered typical for a SN Ic, but high for a SN IcBL \citep{modjaz20}.
Using these narrow lines, a redshift of $z = 0.0037 \pm 0.0002$ is derived and used to de-redshift the spectra. Note that this measurement corresponds to the explosion site, which yields a slightly different value compared to the global host redshift value ($z = 0.00340 \pm 0.00001$; a difference of $\sim90$ km~s$^{-1}$).

\subsection{Light curves and color evolution}

\begin{figure*}
   \centering
   \includegraphics[width=0.8\textwidth]{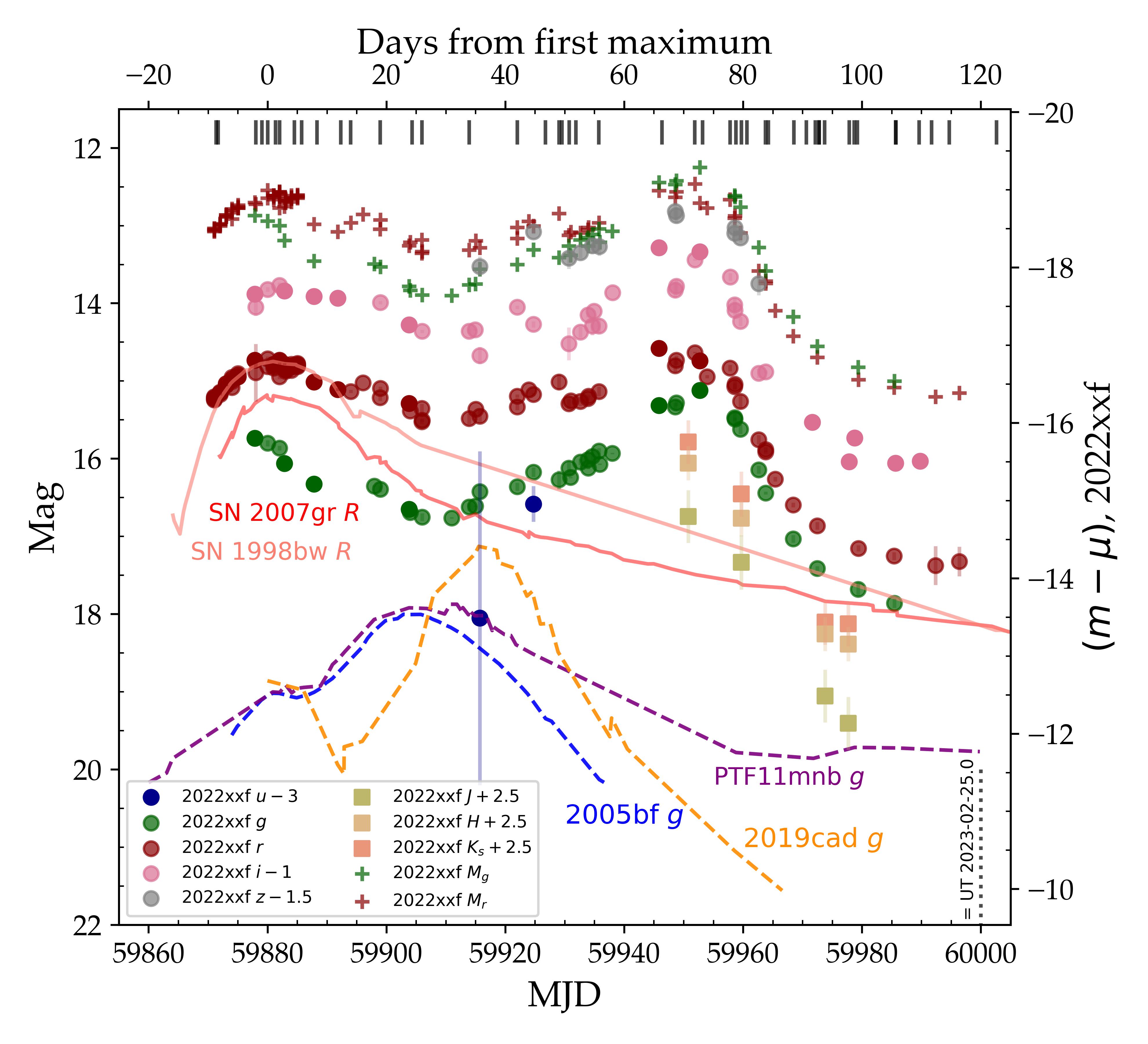} 
   \caption{LCs of SN~2022xxf in \textit{ugrizJHK$_s$} bands. Apparent magnitudes (corrected for Milky Way extinction) are plotted in circles and squares, and dereddened absolute magnitudes ($M_g$ and $M_r$, assuming $E(B-V)_\textrm{total} = 0.8$ mag and $R_V = 3.1$) in plus signs. 
   The LCs of SNe 1998bw, 2007gr, 2005bf, 2019cad, and PTF11mnb are plotted for comparison, shifted in time to match their maximum epochs with the first maximum of SN~2022xxf, and in mag to match their first maxima. Vertical lines on top indicate spectroscopic epochs.}
   \label{fig:lc}%
\end{figure*}

The light curves of SN~2022xxf are presented in Fig.~\ref{fig:lc}. 
Following the discovery, SN~2022xxf rises for more than 10 days to reach a maximum {of} $r = 14.8$ mag. While the very early {rising phase} is not well covered, the LCs around the first peak are consistent with those of typical SNe Ic/IcBL\footnote{Well-observed Type Ic SN~2007gr \citep[e.g.][]{hunter09} and Type IcBL SN~1998bw \citep[e.g.][]{patat01,clocchiatti11} were chosen as representatives. The photometry of these and the other comparison objects were obtained from the Open Astronomy Catalog, \url{https://github.com/astrocatalogs/OACAPI}}.
With \textit{g}-band LC {being} relatively fainter than the \textit{r} and \textit{i} bands, the color appears red, and the \textit{u}-band detections are very faint.
Assuming the host galaxy distance modulus ($\mu = 31.54$ mag) mentioned above and no extinction, the SN magnitude at the first maximum corresponds to an absolute magnitude of $M_r = -16.8$ mag. This is underluminous for a SN IcBL, but still within the observed range for SNe Ic \citep{taddia15,Sollerman2022}. 
{If correcting for a reddening of $E(B-V) = 0.8$ mag} (assuming $R_V = 3.1$), the LCs of SN~2022xxf become brighter, reaching {an} absolute {peak} magnitude of $\sim -19$~mag {in $r$-band}, well within the range of SNe IcBL \citep[e.g.][]{perley20}. 

\begin{figure}
   \centering
   \includegraphics[width=0.5\textwidth]{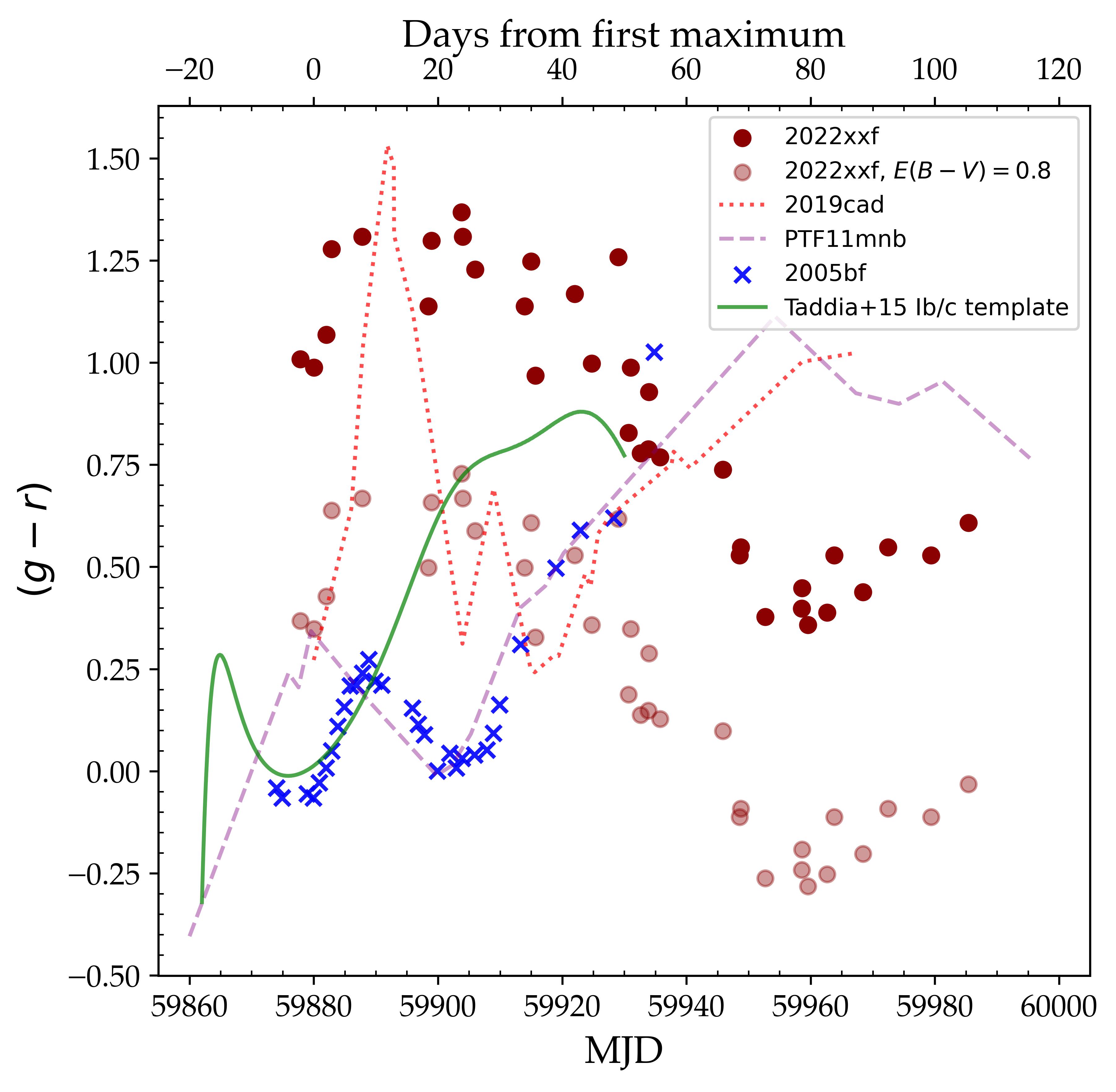} 
   \caption{Color curve of SN~2022xxf in {$(g-r)$}, considering the observed and dereddened cases, as compared to those of SNe 2005bf, 2019cad, and PTF11mnb. Milky Way {reddening is removed from the color curves} in all cases.
   The SN Ib/c color curve template from \citet{taddia15} is also plotted for reference.
   }
   \label{fig:colcurve}%
\end{figure}

The color evolution of SN~2022xxf supports significant extinction. As seen in Fig.~\ref{fig:colcurve}, the dereddened $(g-r)$ color of SN~2022xxf during the first hump is roughly consistent with that of typical SNe Ib/c around and after the main LC peak. {If the color curves are not corrected for reddening, the observed color} of SN~2022xxf would be very red.
An extra $E(g-r)$ of $\sim0.3$ mag would further be required to bring the colors of SN~2022xxf to match the SN Ib/c template of \citet{taddia15}, although given the peculiar nature of SN~2022xxf it does not necessarily have to show the same colors as regular SNe Ib/c.

In SN~2022xxf, following the first maximum (MJD 59880.0 adopted as phase $\phi = 0$~day), the LCs decline and at around +30 days they rise again in all bands. The second rise lasts longer, for about 40 days, after which the SN reaches a second maximum at around +75 days. The second maximum is slightly brighter compared to the first one by $0.1-0.2$ mag in the $r$ and $i$ bands, and clearly brighter (0.6 mag) in the $g$ band.
The \textit{u}-band rise is dramatic ($\sim$0.1 mag~d$^{-1}$), although there are only two data points, {and the first one has a large error bar}.
This implies a color evolution towards the blue in the second LC hump, which is confirmed in the $(g-r)$ color curve (Fig.~\ref{fig:colcurve}).
As the blue turnover is irrespective of the line-of-sight extinction, this is a robust observation.
The peak brightness of the second peak is about 2 mag brighter than the luminosity expected from the $^{56}$Co decay tail, equivalent to {an increase of luminosity by more than six times}.

To estimate the total radiative energy of the second hump, we calculate the bolometric LC of SN~2022xxf (Fig.~\ref{fig:lcbol}) using the method of \citet{lyman14}. {A reddening} of $E(B-V)=0.8$~mag is assumed and corrected.
Assuming that the first hump can be represented by a SN~1998bw-like LC, the bolometric LC of SN~2022xxf is subtracted by that of SN~1998bw, computed in the same way and scaled down to match the first maximum of SN~2022xxf (Fig.~\ref{fig:lcbol}). The difference LC, which represents the second rise and hump, is then integrated along time, resulting in a total radiative energy of $\sim 4 \times 10^{49}$ erg. This amounts to a significant fraction of the total radiative energy ($\sim 7 \times 10^{49}$ erg), but can be achieved in the interaction scenario that requires only a few percent of typical SN explosion energy of $10^{51}$ erg converted into radiation.
Using the Hybrid Analytic Flux FittEr for Transients (HAFFET)\footnote{\url{https://github.com/saberyoung/HAFFET}} tool \citep{yang23}, the first hump could be fit with a standard Ni-powered model with $\sim0.4~M_\odot$ of $^{56}$Ni, while the slow rise of the second hump prevents a reliable fit with $^{56}$Ni.

The second hump of the LC drops faster than it rises, which suggests that $^{56}$Ni heating is unlikely to be the cause of the second hump; the decline is usually slower than the rise in the $^{56}$Ni heating mechanism (e.g., see SNe 1998bw and 2007gr in Fig.~\ref{fig:lc}). 
The production of $^{56}$Ni requires high temperature and density, thus this could only occur deep in the core, which implies that it is unlikely for the SN to synthesize any significant amount of $^{56}$Ni well after the explosion.
Due to the longer photon diffusion time as the ejecta expand, $^{56}$Ni heating would predict a decline rate that is slower than the rise, which does not match the observations of SN 2022xxf. 
Therefore, a different mechanism is likely to be at play.
After the fall, the LC settles at the level of the tail luminosities of SNe 1998bw and 2007gr (after being scaled at the first peak luminosity), indicating that the first peak is powered by the canonical $^{56}$Ni heating. There is however a hint of the LC flattening in SN~2022xxf ($>$ +100 days), which suggests that extra radiative energy is still generated in SN~2022xxf in addition to the power from the radioactive decay input. 

{A few well-observed two-humped SESNe in the literature are also overplotted in Fig.~\ref{fig:lc}, including} SN~2005bf \citep{anupama05,tominaga05,folatelli06}, its analogs PTF11mnb \citep{taddia18} and SN~2019cad \citep{gutierrez21}. {The} LC of SN~2022xxf appears to be distinct, although there is a diversity in the LCs of the other objects as well. 
{For those comparison SNe}, the second LC peak occurs $20-30$ days after the first peak, i.e. in less than half the time compared to SN~2022xxf. 
{After} the second maximum, the LCs of the {comparison} objects decline slowly, whereas in the case of SN~2022xxf a more sudden drop in all bands follows ($\sim0.1$ mag~d$^{-1}$). As for the peak magnitudes, these other two-humped SNe peak at $\sim -18$~mag\footnote{SN~2019cad could have reached nearly $-20$~mag if corrected for $E(B-V) = 0.49$ mag \citep{gutierrez21}.} at the second maximum, while SN~2022xxf is potentially brighter at $\sim -19$~mag, though we note the uncertainties in both distance estimates and in the extinction corrections. 

All these objects {including SN~2022xxf} show a blue turnover in the color evolution during the second rise, after which they become redder again. 
Only when the SNe approach the second hump do the colors become bluer than expected. This behavior may be interpreted as a `normal' radioactive first peak, followed by a second peak possibly powered by another mechanism generating extra energy. Several explanations for the powering of the second peak have been offered: an asymmetric explosion \citep{folatelli06}, a bimodal nickel distribution \citep{tominaga05,taddia18,gutierrez21}, or magnetar power \citep{maeda07,gutierrez21}. CSM interaction has not been invoked as a possible mechanism, due to the lack of strong narrow spectral lines, although it has been suggested that CSM interaction does not always require the production of narrow emission features \citep[e.g.][]{chugai01,sollerman20,dessart22,maeda2023}.
The color evolution of SN~2022xxf (Fig.~\ref{fig:colcurve}) suggests that it becomes bluer as it approaches the second peak, starting from around $+30$ days. At the second peak, it has become bluer than the first peak by $(g-r)\sim0.7$~mag, and subsequently the color becomes redder again as it falls from the peak. A similar blue turnover is seen in the other double-hump objects, although none of them reached colors bluer than $(g-r) = 0$~mag as in the case of SN~2022xxf, and they \emph{never} became bluer than {the observed colors seen} during the first peak. Such very blue colors are seen in SESNe interacting with H-poor CSM, such as SNe Ibn \citep[e.g.][their Fig.~12]{ho21}, Icn \citep{galyam22,pellegrino22}, Type Ic SN~2010mb \citep{benami14} and SN~2021ocs \citep{hk22}.

\subsection{Optical spectra}

\begin{figure*}
   \centering
   \includegraphics[width=\textwidth]{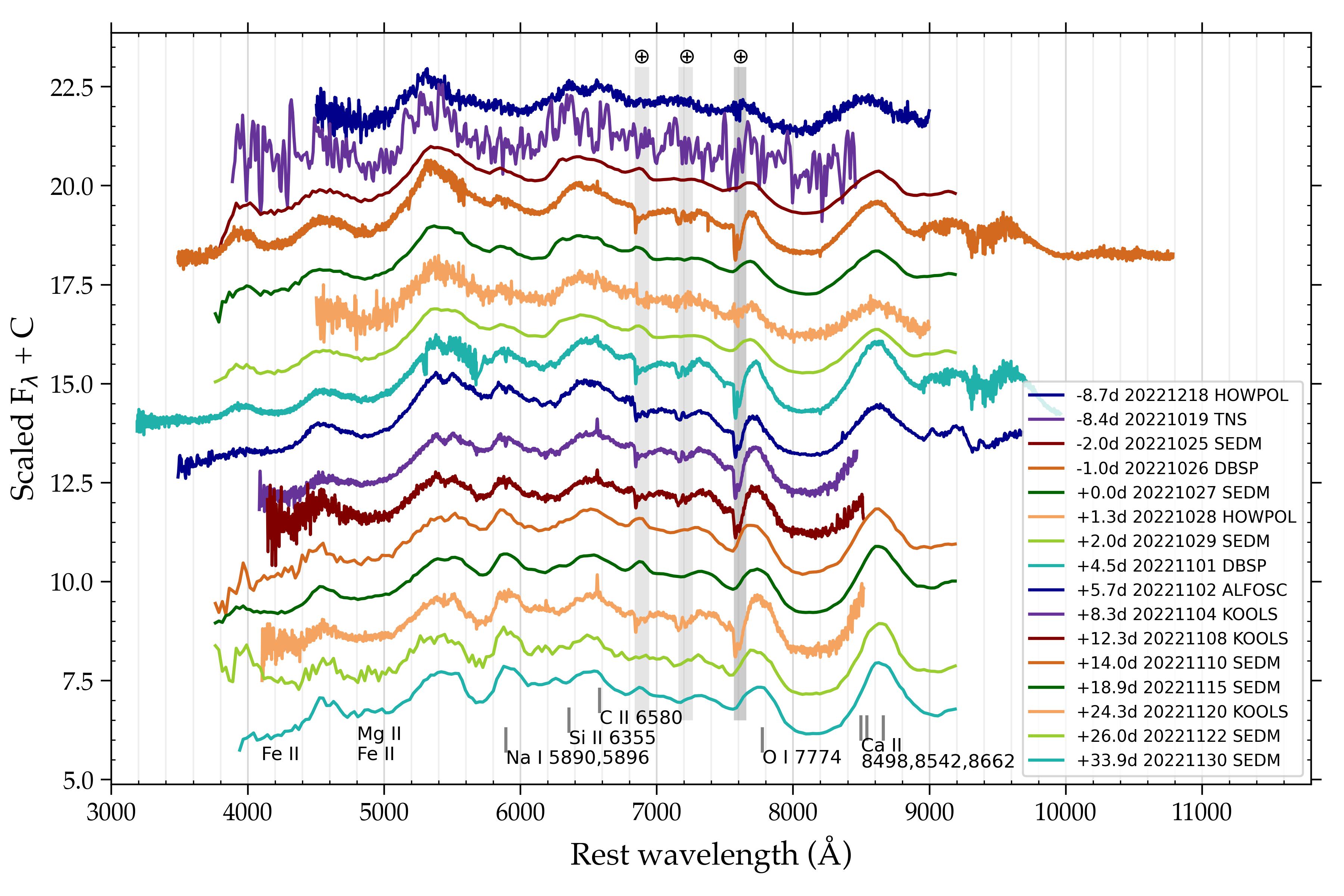} 
   \caption{Spectral sequence of SN~2022xxf during the first light-curve hump. Spectra are normalized by their average values and not corrected for reddening. 
   Phases are in observer frame, relative to the first LC maximum.
   Prominent spectral lines are indicated with vertical lines corresponding to the rest wavelengths. Vertical grey shades indicate spectral regions affected by telluric absorption.}
   \label{fig:spec1}%
\end{figure*}

\begin{figure*}
   \centering
   \includegraphics[width=\textwidth]{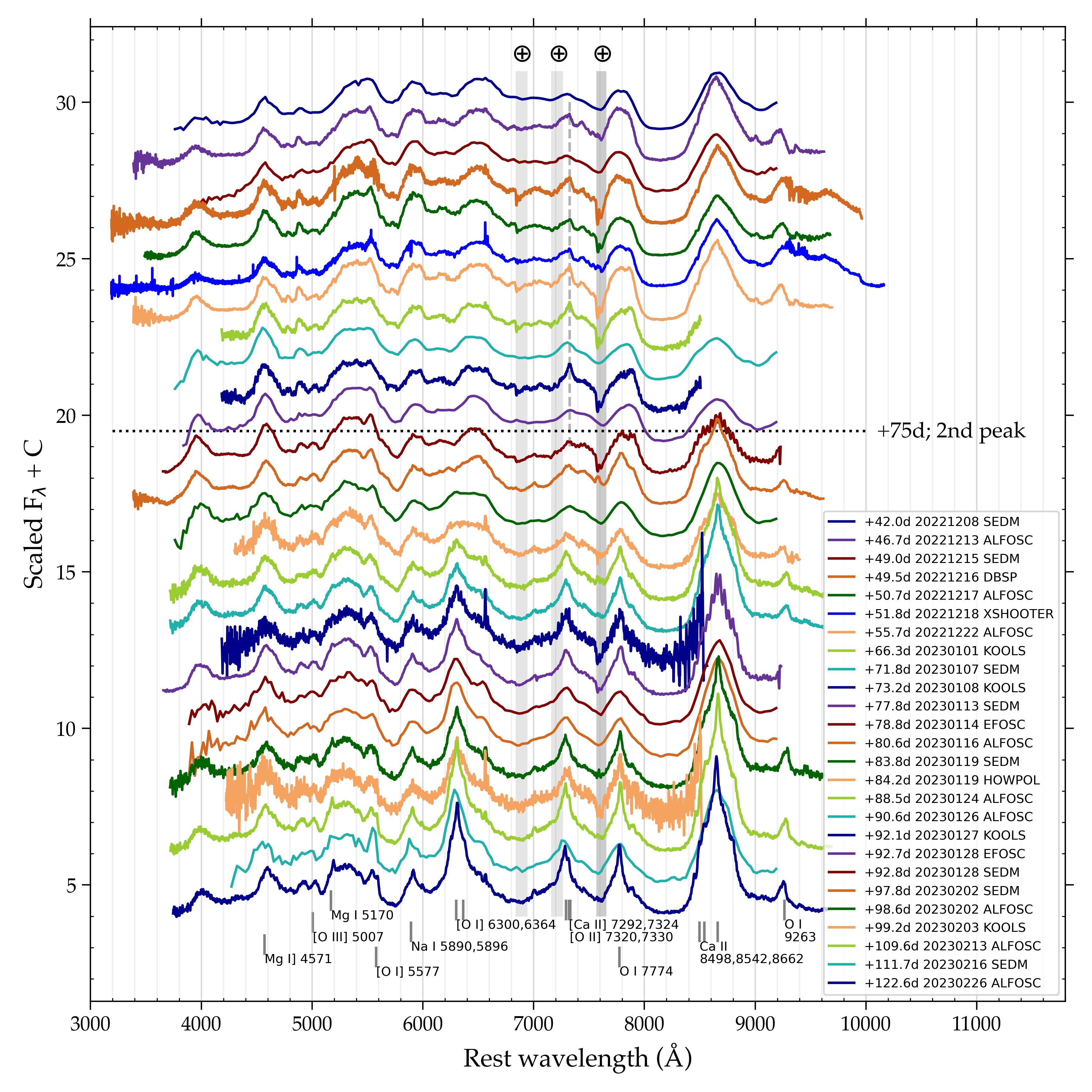} 
   \caption{Spectral sequence of SN~2022xxf during the second light-curve hump. Spectra are normalized by their average values and not corrected for reddening. 
   Phases are in observer frame, relative to the first LC maximum.
   Prominent emission lines are indicated with vertical lines corresponding to the rest wavelengths.
   Dashed vertical line indicates 7325 \AA, the average wavelength of the [O~II] $\lambda\lambda$7320,7330 doublet (see sect. \ref{sec:narrowlines}), and dotted horizontal line the approximate epoch of the second LC peak.
   Vertical grey shades indicate spectral regions affected by telluric absorption.}
   \label{fig:spec2}
\end{figure*}

Figs.~\ref{fig:spec1} and \ref{fig:spec2} show the spectral evolution of SN~2022xxf during the first and second hump, respectively. With nearly 50 high signal-to-noise ratio spectra {collected} within 130 days, this dataset comprises one of the most well-sampled observations of a (SE)SN to date. 
The spectrum initially appears smooth with broad profiles, typical for the early phase, and develops to show more features with time. 
The ejecta velocity {estimated from the absorption minimum of O~I $\lambda$7774 in the first pre-maximum phase spectrum is $\sim14000$ km~s$^{-1}$.}
Compared to some SNe IcBL such as SN~1998bw \citep[e.g.][]{patat01}, this {velocity} is relatively low, but still well within the range of the SN IcBL sample (e.g. \citealt{modjaz16}, their Fig.~5).
The strongest spectral features during the first hump are the P-Cygni profiles of Na~I $\lambda\lambda$5890,5896, Si~II $\lambda$6355, O~I $\lambda$7774, Ca~II $\lambda\lambda$8498,8542,8662 and the deep Fe absorptions at $\sim$ 4200 and 5000 \AA, typically seen in Type Ic/IcBL SNe around peak brightness. This supports the notion that the first LC hump was powered similarly as the LCs seen in the majority of SESNe, via radioactive decay power and not by e.g. a shock cooling mechanism.
During the second hump (Fig.~\ref{fig:spec2}), the spectrum gradually morphs into becoming more nebular, as expected for SESNe a few months post-LC peak \citep[e.g.][]{patat01,hunter09}.  

The overall spectral evolution of SN~2022xxf appears to be gradual, with no abrupt changes, and again similar to that of regular SNe Ic/IcBL (see Fig.~\ref{fig:speccomp}, left). 
The transition from the photospheric to the nebular phase occurs relatively slowly, as in SNe IcBL, with the emergence of the nebular [O~I] $\lambda\lambda$6300,6364 line at around +80 days. In the case of SNe Ic, the emergence of this line could occur earlier at around +60 days (Fig.~\ref{fig:speccomp}) or even before.
There is no sign of H or He emission lines appearing during the second hump, which is otherwise expected in the case of SN ejecta interacting with dense H/He-rich CSM (e.g. SNe 2017dio, \citealt{hk18a}, 2017ens, \citealt{chen18}, 2019oys, \citealt{sollerman20}, and SNe Ibn, e.g. \citealt{pastorello07}). 
When {a reddening of} $E(B-V) = 0.8$ mag is considered, the dereddened spectra of SN~2022xxf appear to have an enhanced brightness at bluer wavelengths, relative to SNe Ic/IcBL at the corresponding epochs (Fig.~\ref{fig:speccomp}, left, {spectra at $\gtrsim 50$ days}). This seems to support the CSM interaction interpretation. The rising blue continuum and strong Fe bump at around 5300 \AA~ are also frequently seen in the late-time spectra of interacting SNe, both in H-rich events such as SNe 2017dio and Type IIn SNe (e.g. SN~2020uem, \citealt{uno23}), and in H-poor events such as SNe Ibn (e.g. SN~2006jc, \citealt{pastorello07}) and Icn (e.g. SN~2019hgp, \citealt{galyam22}). 

Looking at the spectra of the other two-humped objects (Fig.~\ref{fig:speccomp}, right), they and SN~2022xxf share some similarities during the respective LC phases, while clearly there are also differences. We note that the spectral classifications of these objects are not identical: SN~2005bf was thought to be a Type Ic or Ib (or even possibly IIb with the interpretation of some features as hydrogen, \citealt{anupama05}), PTF11mnb and SN~2019cad are both Type Ic SNe with lines narrower compared to those in SN~2022xxf. While the comparison is done for the same parts of the LC anatomy (first peak, valley, second peak, drop), the timescales are different for these objects.
It is therefore likely that the variations seen in the spectra reflect the different time phases, and the emergence and disappearance of the second LC hump do not leave clear traces in the spectral evolution as previously pointed out \citep{folatelli06,taddia18,gutierrez21}. 
{SN~2019cad appears to be the `bactrian' object most similar to SN~2022xxf. While its second LC hump appears earlier, the color evolution and general spectral appearance are relatively similar to those of SN~2022xxf (Figs.~\ref{fig:colcurve} and \ref{fig:speccomp}, right). A possible narrow O~I $\lambda$7774 emission line is seen in the +88.1 day spectrum of SN~2019cad \citep[see Fig. 3 of][]{gutierrez21}, although upon closer examination this feature is most likely noise.}
In {SNe 2005bf, 2019cad, and PTF11mnb}, the LCs could be modeled with a bimodal Ni distribution \citep{orellana22} or with additional magnetar power input \citep{maeda07,gutierrez21}, although the effects of these mechanisms on the spectra are yet to be evaluated.

\subsubsection{Narrow emission lines}
\label{sec:narrowlines}

In addition to the typical broad spectral features, a possible narrow feature that is likely related to the second LC hump is identified in SN~2022xxf.
During the rise to the second peak (Fig.~\ref{fig:spec2}, from around day +46 to +73), the spectra of SN~2022xxf show a weak emission line at 7325~\AA, with a velocity FWHM of initially $\sim 4500$ km~s$^{-1}$  which becomes narrower and reaches $\sim 2500$ km~s$^{-1}$. This line then gradually disappears after the second peak, resulting in a flatter profile in that spectral region, before eventually the nebular [Ca II] $\lambda\lambda$7292,7324 line starts to appear and take over (Fig.~\ref{fig:line7325}). The peak of the transient line at 7325~\AA~ is close to the $\lambda$7324 component of the [Ca II] doublet, although this is not accompanied by the $\lambda$7292 component. It is redder than the average wavelength of the [Ca II] doublet, and thus may be better associated with the [O II] $\lambda\lambda$7320,7330 line.

\begin{figure*}
   \centering
   \includegraphics[width=\textwidth]{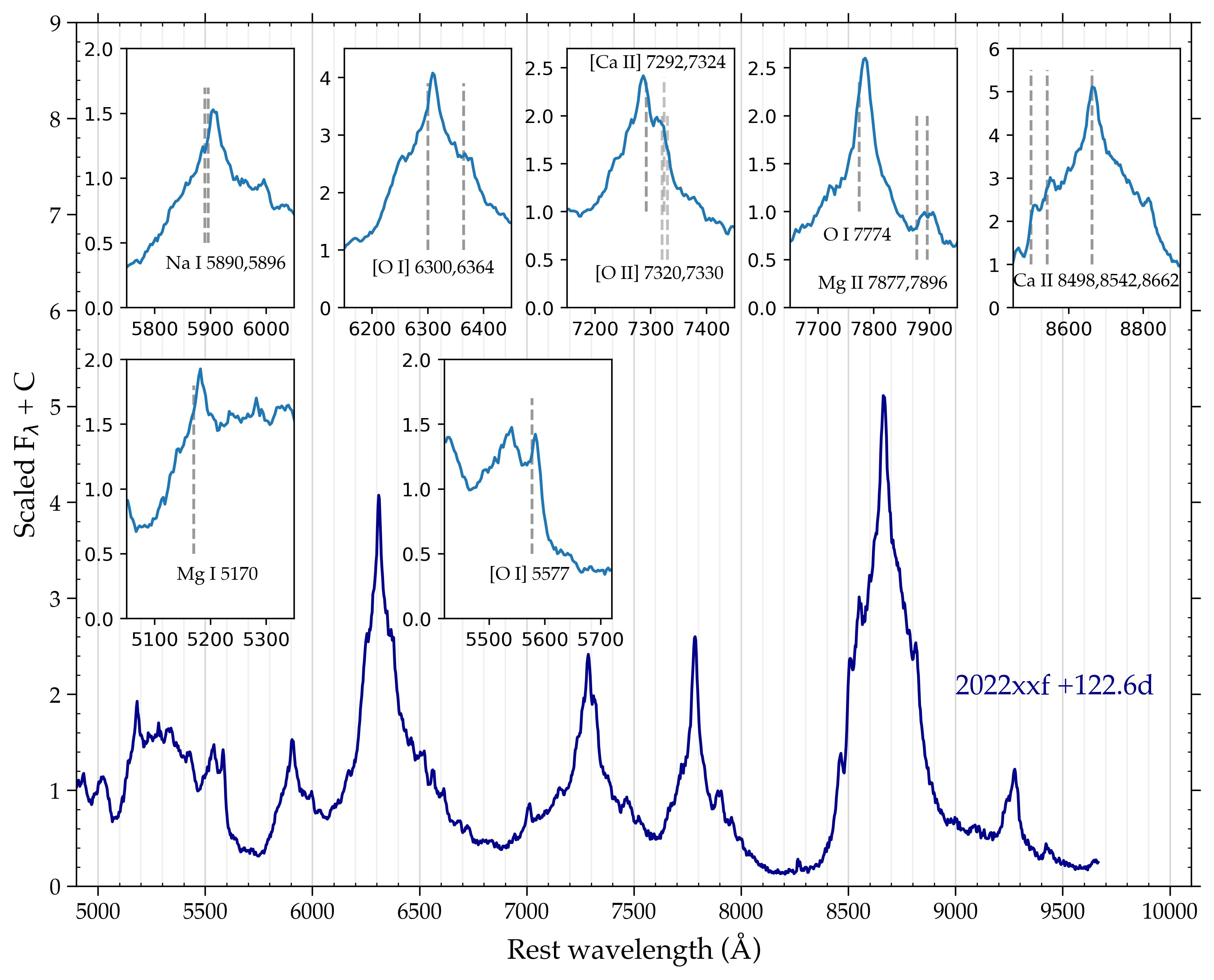} 
   \caption{Nebular spectrum of SN~2022xxf at +122.6 days. Insets show close-ups of the emission lines, with dashed vertical lines indicating the rest wavelengths of the emitting species.}
   \label{fig:nebspec}
\end{figure*}

Narrow emission lines are more easily seen in the nebular phase of SN~2022xxf. Following the second maximum, narrow features emerge, superposed on the broad emission lines. All the broad nebular lines show a narrow core profile, indicating emission from low velocities. This narrow line profile is seen in Mg~I $\lambda$5170, [O~I] $\lambda$5577, and all the major lines redwards i.e. Na~I $\lambda$5893, [O~I] $\lambda\lambda$6300,6364, [Ca~II] $\lambda\lambda$7292,7324, O~I $\lambda7774$, O~I $\lambda8446$, the Ca~II triplet (all the individual triplet components), O~I $\lambda$9263, and other smaller structures across the spectrum (as shown in Fig.~\ref{fig:nebspec}). 
These {narrow features} are unlikely to originate from the host galaxy.
The narrow components are typically narrower than 2500 km~s$^{-1}$ (FWHM), superposed on broad components of $\sim5000$ km~s$^{-1}$. The peaks are generally offset by +$300-500$~km~s$^{-1}$ from zero velocity inferred from the redshift, though this is comparable to the instrument resolutions in the case of the low-resolution spectroscopy.
The narrowest emission lines such as Mg~I $\lambda$5170 and [O~I] $\lambda$5577 display velocities up to $\pm 500-800$~km~s$^{-1}$ at their bases (HWZI, half-width at zero intensity), suggesting that they are unresolved given the instrument resolution\footnote{On the other hand, these velocities could be underestimated as the lines are situated on a pseudocontinuum, and thus their zero flux levels are uncertain.}. 
In addition, other, mostly weaker, narrow lines are also found, including both known and unidentified lines, at 5532, 7010, 7155 ([Fe II]), 7470, 7877/7896 (Mg II doublet), 8270, 8815, and 9436 (Mg I) \AA, some of which are seen already in the early nebular phase shortly after the LC drop, or even during the second rise in the case of 5532 \AA~ (Fig.~\ref{fig:spec2}). The [O III] $\lambda$5007 line is variable and relatively broad, $\sim \pm$3000 km~s$^{-1}$ at the base.
In comparison, SNe Icn display narrow C/O emission lines corresponding to velocities $1000-2000$~km~s$^{-1}$ \citep{galyam22,perley22}, and SNe Ibn show $\sim$ 2000 km~s$^{-1}$ in the He~I emission lines \citep{pastorello07}.
We do not detect high-ionization coronal lines \citep[see e.g.][]{fransson14,chen18,sollerman20} in the intermediate-resolution X-Shooter spectrum, although it is likely that the CSM interaction was still weak at +50 days.
The line profiles are generally Gaussian, without {showing} the wings of a Lorentzian profile, indicating that electron scattering is not significant. This is expected in ejecta dominated by intermediate mass elements, e.g. plasma of singly-ionized O will have 1 electron per 16 nucleon, therefore the electron scattering opacity in cm$^2$~g$^{-1}$ is 16 times lower than that of a H-rich plasma (1 electron per 1 proton).

\subsection{Near-infrared spectra}

\begin{figure*}
\centering
\begin{subfigure}{.5\textwidth}
  \centering
  \includegraphics[width=\linewidth]{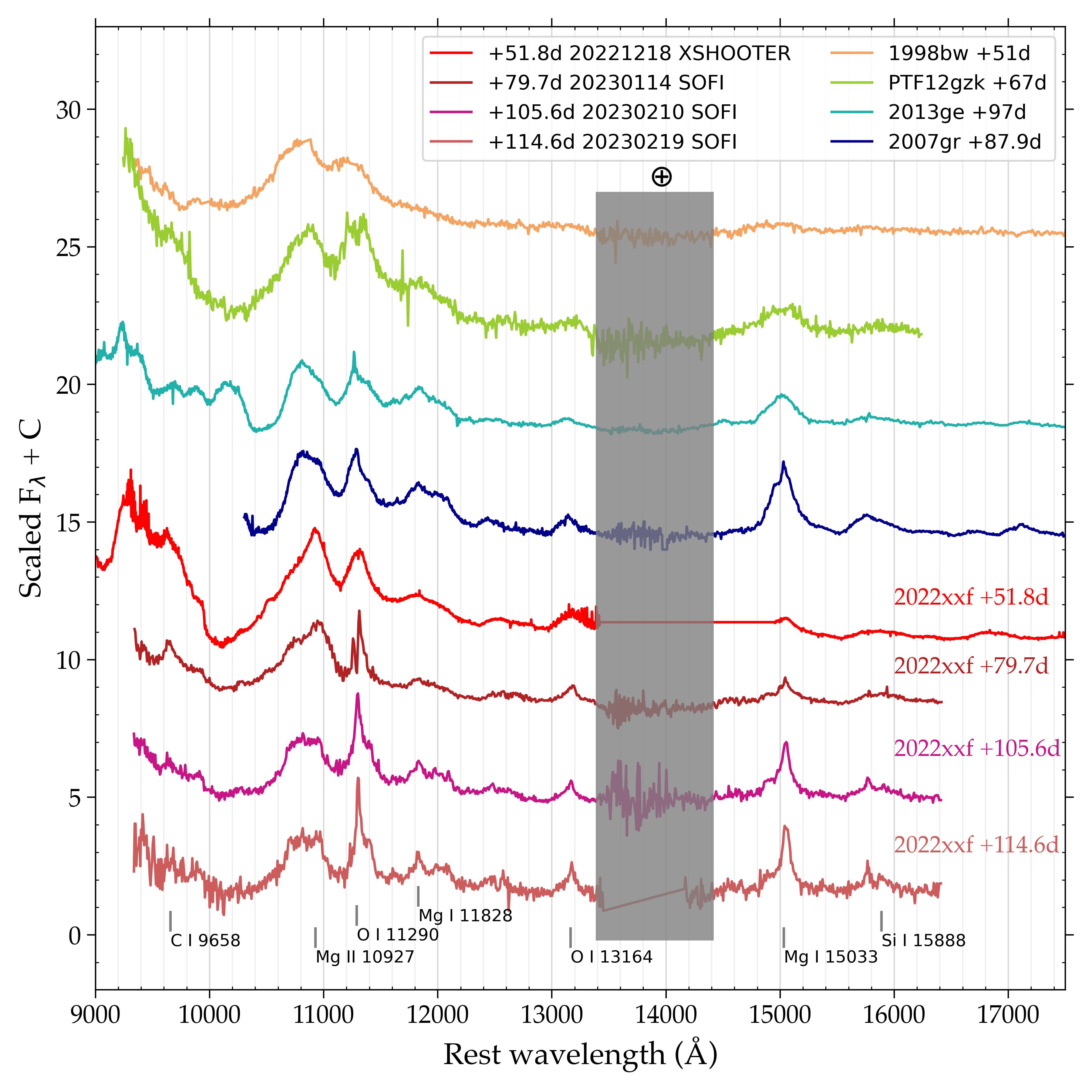}
\end{subfigure}%
\begin{subfigure}{.5\textwidth}
  \centering
  \includegraphics[width=\linewidth]{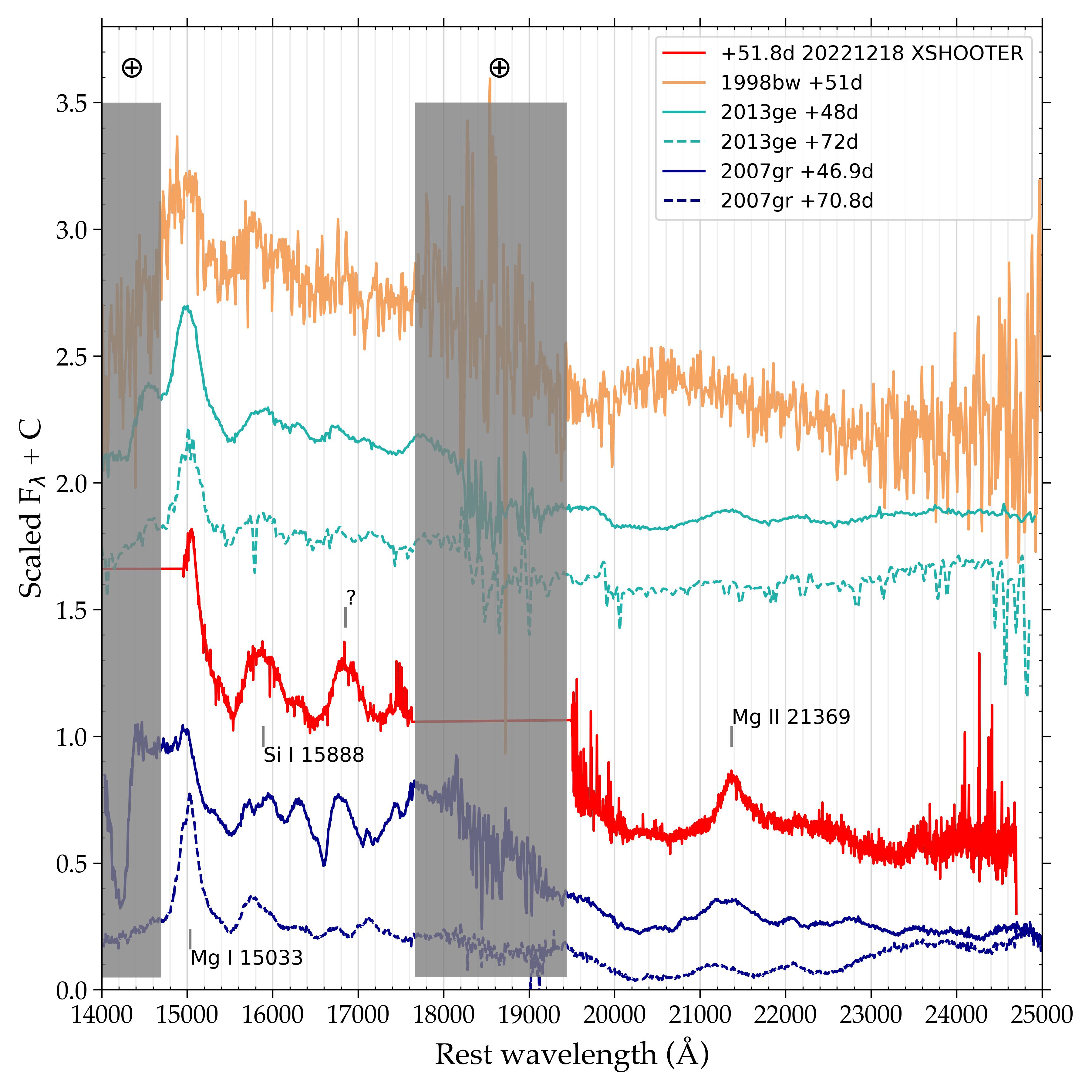}
\end{subfigure}
\caption{NIR spectra of SN~2022xxf in the \textit{JH} (\textit{left panel}) and \textit{HK} bands (\textit{right panel}), compared to those of SN IcBL 1998bw \citep{patat01} and SNe Ic PTF12gzk (\citealt{horesh13}; spectrum from PESSTO Data Release 1, \citealt{Smartt2015}), 2013ge \citep{drout16}, 2007gr \citep{hunter09}. Prominent emission lines are indicated with vertical lines corresponding to the rest wavelengths. Vertical grey shades indicate spectral regions affected by telluric absorption.
\label{fig:specnir}}
\end{figure*}

The near-infrared (NIR) spectra of SN~2022xxf are presented in Fig.~\ref{fig:specnir}. 
Here, we follow the NIR line identifications of \citet{hunter09,rho21,shahbandeh22}. 
Relative to the optical spectrum, the NIR spectrum becomes nebular earlier due to the lower optical depth at longer wavelengths. In comparison to other Type Ic/IcBL objects, while the rarity of NIR spectra of these objects especially at later phases only allows for limited comparisons, the earliest spectrum of SN~2022xxf at +51.8 days from X-Shooter displays a similar global appearance with broad emission lines near 10800, 11200, 11800, 13200, 15000, and 16000 \AA.
The main difference is the narrow peak of the Mg II $\lambda$21369 emission line, which is not seen in the other objects, and a broad unidentified line at $\sim16850$ \AA~with similar strength as Si I $\lambda$15888.
The subsequent spectra show a similar set of features, as seen also in other SNe Ic/IcBL although previous observations rarely reach later than +100 days.
As in the optical, there are no clear detections of H and He lines. The feature at 10800 \AA~ is attributed to a blend of C~I/Mg~II/O~I and may contain He~I 10830 \AA, though not accompanied by He~I 20581 \AA, as in the case of SNe Ic/IcBL \citep[][e.g. their Fig.~5]{shahbandeh22}.

As the evolution progresses, the emission lines of SN~2022xxf gradually show narrower profiles (see Fig.~\ref{fig:lineprofs}). O I $\lambda$11290 is clearly showing a sharp peak, and curiously a split line profile in the +79.7 days spectrum. Other lines also show broad to narrow evolution, e.g. Mg I $\lambda$11828 and $\lambda$15033, as well as O I $\lambda$13164. The broad profile at 10800 \AA~initially shows a pronounced peak consistent with Mg II $\lambda$10927, which slowly decays resulting in a flat-topped profile at later epochs. 
Following the LC drop ($>$ +75 days), the Mg~I lines become stronger relative to the O~I lines. Mg II and C I also weaken during the same time period, which suggests cooling evidenced by the growing Mg I. 
Si I $\lambda$15888 appears initially as a broad profile before developing a narrow peak, accompanied by an unidentified stronger peak bluewards at 15770 \AA. The narrow features in the NIR spectra of SN~2022xxf are not seen in the other SNe at the corresponding epochs. They are coeval with those seen in the optical and show similar velocities, evolving from $\sim4000-5000$ to $\sim$2000 km~s$^{-1}$.

\section{CSM and progenitor properties}

{The rebrightening of the LC, blue color evolution, and narrow emission line profiles suggest that ejecta-CSM interaction is likely to be significant in SN~2022xxf.
While it is unlikely that the immediate vicinity of the progenitor star was completely free of CSM, and thus ejecta-CSM interaction could have taken place also in the early phases although relatively weakly, the second LC hump is naturally explained by interaction with the bulk of the CSM.
If the LC rise starting around 45 days after explosion (assuming that the explosion occurred  $\sim$ $15$ days before the first maximum) corresponds to the inner edge of a detached CSM, then with a $\sim$ 14000 km~s$^{-1}$ ejecta expansion velocity the location of the inner edge of the CSM would be around $5 \times 10^{15}$ cm. 
If such CSM were formed by material ejected from the progenitor star at $\sim2000$ km~s$^{-1}$, this ejection must have occurred within $\sim 1$ year prior to the explosion.
It is to be noted that the velocities seen in the narrow lines may reflect a combination of expansion velocity of an unshocked CSM and the shock velocity in the ejecta-CSM interaction, therefore both velocity components are in reality slower than the line width implies, which results in results in a longer lag time between the pre-SN mass ejection and the SN explosion.  
}

{In the literature, it has been argued that mass-loss episodes within short pre-SN timescale could be caused by e.g. wave-driven outbursts \citep{wu21}, centrifugally-driven mass loss through spin-up \citep{aguilera18}, or pair-instability pulsations \citep{renzo20}.
Observationally, pre-SN outbursts on such timescales have been reported for the Type Ibn SNe 2006jc and 2019uo \citep{pastorello07,strotjohann21}.
For SN~2022xxf, we searched ZTF data up to 4.8 years prior to the SN with a total time coverage of 11\% during this period, but no precursor eruption was found down to absolute magnitude of $\sim -11$ mag.
}

{SN~2022xxf may be related to the emerging subclass of SNe Ic with H/He-poor CSM interaction (`Ic-CSM'), such as SNe 2010mb \citep{benami14} and 2021ocs \citep{hk22}. These objects also show O and Mg lines in their spectra (Fig.~\ref{fig:compneb}), slow-evolving unusual LCs, and very blue colors. Their spectral evolutions are not well sampled, unfortunately, although SN~2010mb shows persistent nebular lines and a narrow [O I] $\lambda$5577 emission line which is also seen in SN~2022xxf.
The \textit{r}-band LC of SN~2010mb shows a bumpy 180-days `plateau' at $M_r \approx -18.3$ mag. 
It is, however, unclear how these SNe are related to the Type Icn SN class \citep{galyam22,pellegrino22} although they all share the property of interaction with H/He-deficient CSM. The Ic-CSM objects show relatively regular photospheric to nebular spectral evolution and long-lasting LCs, in stark contrast to the current sample of SNe Icn. This suggests that the properties of the progenitors and the CSM could be very disparate. {The distribution of the CSM must be different, pointing to different mass-loss episodes; SNe Icn show a confined CSM rapidly decreasing outward \citep{nagao23}, similar to the case for SNe Ibn \citep{maeda2022}, while the SNe Ic-CSM objects should have a more extended CSM.} 
It is likely that the masses of ejecta and C/O/Mg CSM in SNe Icn and Ic-CSM are considerably different, where it could be low in the former and high in the latter ($\sim 3~M_\odot$ of CSM and $\gtrsim 10~M_\odot$ of ejecta in the case of SN~2010mb, \citealt{benami14}).
An explanation has recently been offered by \citet{tsuna23}, in which the progenitors of both SNe Ibn/Icn and Ib/c-CSM similarly form the CSM by pre-SN mass ejection. The differences in the CSM fallback process and SN explosion timing naturally explain the different observational properties between the two subclasses.
In this picture, the C-O star progenitor of SN~2022xxf could have experienced a mass ejection with a weak fallback (which is regulated by the interplay of the infalling material and the radiation pressure of the star), resulting in a detached CSM configuration.
}

\section{Summary and conclusions}
\label{sec:conclusion}

{SN~2022xxf displays an unprecedented LC evolution with two distinctive humps separated by $\sim75$ days and a $\sim2$ mag peak-to-valley amplitude, suggesting an extra energy input on the order of $4\times10^{49}$ erg in addition to the regular $^{56}$Ni decay powering the first hump. The global optical/NIR spectral evolution is similar to the population of SNe Ic/IcBL, although the emergence of narrow features in SN~2022xxf during and after the second LC hump suggests the presence of a slower-moving material at $1000-2000$~km~s$^{-1}$. If this were due to CSM, the second hump may be explained by ejecta-CSM interaction producing extra radiative energy from the conversion of kinetic energy, although this did not affect the spectra significantly as the narrow emission lines are subtle. 
A CSM interaction scenario is also supported by the dramatic blue color evolution, the slow rise and fast drop of the LC in the second hump, and the flattening of the tail phase. 
SN~2022xxf thus represents another rare example of a H/He-poor SN interacting with a H/He-poor CSM.
}

{The properties of the CSM, ejecta, and progenitor star of SN~2022xxf are subject to further study involving long-term monitoring and multi-wavelength observations (Izzo et al., in preparation). 
Observations at later epochs may reveal additional clues on the origin and composition of the CSM, and therefore the associated mass loss of the progenitor star before the explosion.}

\begin{acknowledgements}
All the spectral data of comparison objects were obtained from the WISeREP repository \citep[][\url{https://www.wiserep.org/}]{yaron12}, where the data of SN~2022xxf will be published as well.
{The anonymous referee and}
Schuyler Van Dyk {are} thanked for helpful suggestions on the manuscript.
We thank the following for obtaining some of the observations: Takashi Nagao, William Meynardie, Yu-Jing Qin, Shreya Anand, Tomas Ahumada, Jean Somalwar, Kaustav Das, Miranda Kong.
H.K. was funded by the Academy of Finland projects 324504, 328898, and 353019.
K.M. acknowledges support from the JSPS KAKENHI grant Nos. JP18H05223, JP20H00174, and JP20H04737. K.M also acknowledges Koichi Itagaki for his private notice on the discovery of SN 2022xxf immediately after the TNS report. 
MWC is supported by the National Science Foundation with grant numbers PHY-2010970 and OAC-2117997.
MMK acknowledges generous support from the David and Lucille Packard Foundation.
MG is supported by the EU Horizon 2020 research and innovation programme under grant agreement No 101004719.
CA acknowledges support by NASA grant JWST-GO-02114.032-A and JWST-GO-02122.032-A.
MN is supported by the European Research Council (ERC) under the European Union’s Horizon 2020 research and innovation programme (grant agreement No.~948381) and by funding from the UK Space Agency.
PL acknowledges support from the Swedish Research Council.
L.G. and T.E.M.B. acknowledge financial support from the Spanish
Ministerio de Ciencia e Innovaci\'on (MCIN), the Agencia Estatal de Investigaci\'on (AEI) 10.13039/501100011033, the European Social Fund (ESF) "Investing in your future", and the European Union Next Generation EU/PRTR funds under the PID2020-115253GA-I00 HOSTFLOWS project, the 2019 Ram\'on y Cajal program RYC2019-027683-I, the 2021 Juan de la Cierva program FJC2021-047124-I, and from Centro Superior de Investigaciones Cient\'ificas (CSIC) under the PIE project 20215AT016, and the program Unidad de Excelencia Mar\'ia de Maeztu CEX2020-001058-M.
SM acknowledges support from the Magnus Ehrnrooth Foundation and the Vilho, Yrj\"{o}, and Kalle V\"{a}is\"{a}l\"{a} Foundation.
YZC is supported by International Centre of Supernovae, Yunnan Key Laboratory (No. 202302AN360001).
P.C. acknowledges support via an Academy of Finland grant (340613; P.I. R. Kotak).
Q.F. acknowledges support by JSPS KAKENHI Grant (20J23342).
This research was supported by the Munich Institute for Astro-, Particle and BioPhysics (MIAPbP) which is funded by the Deutsche Forschungsgemeinschaft (DFG, German Research Foundation) under Germany's Excellence Strategy –- EXC-2094 -– 390783311.
The work is partly supported by the JSPS Open Partnership Bilateral Joint Research Project between Japan and Finland (JPJSBP120229923) and also between Japan and Chile (JPJSBP120209937). 
This work was supported by grants from VILLUM FONDEN (project number 16599 and 25501). 
Based in part on observations obtained with the Samuel Oschin Telescope 48-inch and the 60-inch Telescope at the Palomar Observatory as part of the Zwicky Transient Facility project. ZTF is supported by the National Science Foundation under Grants No. AST-1440341 and AST-2034437 and a collaboration including current partners Caltech, IPAC, the Weizmann Institute of Science, the Oskar Klein Center at Stockholm University, the University of Maryland, Deutsches Elektronen-Synchrotron and Humboldt University, the TANGO Consortium of Taiwan, the University of Wisconsin at Milwaukee, Trinity College Dublin, Lawrence Livermore National Laboratories, IN2P3, University of Warwick, Ruhr University Bochum, Northwestern University and former partners the University of Washington, Los Alamos National Laboratories, and Lawrence Berkeley National Laboratories. Operations are conducted by COO, IPAC, and UW.
SED Machine is based upon work supported by the National Science Foundation under Grant No. 1106171.
Based in part on observations made with the Nordic Optical Telescope, owned in collaboration by the University of Turku and Aarhus University, and operated jointly by Aarhus University, the University of Turku and the University of Oslo, representing Denmark, Finland and Norway, the University of Iceland and Stockholm University at the Observatorio del Roque de los Muchachos, La Palma, Spain, of the Instituto de Astrofisica de Canarias, under programmes
66-506 (PI: Kankare, Stritzinger, Lundqvist), 64-501 (PI: Sollerman, Goobar), and 62-507 (PI: Angus).
The data presented here were obtained in part with ALFOSC, which is provided by the Instituto de Astrofisica de Andalucia (IAA) under a joint agreement with the University of Copenhagen and NOT.
The ZTF forced-photometry service was funded under the Heising-Simons Foundation grant $\#$12540303 (PI: Graham). 
The data from the Seimei and Kanata telescopes were obtained under the KASTOR (Kanata And Seimei Transient Observation Regime) project, specifically under the following programs for the Seimei Telescope at the Okayama observatory of Kyoto University (22B-N-CT10, 22B-K-0003, 23A-N-CT10, 23A-K-0006). The Seimei telescope is jointly operated by Kyoto University and the Astronomical Observatory of Japan (NAOJ), with assistance provided by the Optical and Near-Infrared Astronomy Inter-University Cooperation Program. The authors thank the TriCCS developer team (which has been supported by the JSPS KAKENHI grant Nos. JP18H05223, JP20H00174, and JP20H04736, and by NAOJ Joint Development Research). 
Based on observations collected at the European Organisation for Astronomical Research in the Southern Hemisphere, Chile, as part of ePESSTO+ (the advanced Public ESO Spectroscopic Survey for Transient Objects Survey).
ePESSTO+ observations were obtained under ESO program IDs 106.216C and 108.220C (PI: Inserra).
This work was funded by ANID, Millennium Science Initiative, ICN12\_009.
The Aarhus supernova group is funded by the Independent Research Fund Denmark (IRFD, grant number 10.46540/2032-00022B).
\end{acknowledgements}

\begin{appendix} 

\section{Additional table and figures}

\begin{table*}[h!]
\caption{\label{tab:instru}Facilities used in the observations.}
\centering
\begin{tabular}{lcccl}
\hline\hline
Telescope & Instrument & Band (\AA) & $R = \lambda/\Delta\lambda$ & Reference \\
\hline
Kanata & HOWPol & 4500–9000  & 400 & \citet{Kawabata2008} \\
NOT & ALFOSC & 3500-9500  & 400 & URL\tablefootmark{1} \\
NTT & EFOSC2 & 3600-9200  & 400 & \citet{buzzoni84} \\
NTT & SOFI & 9300-16400  & 600 & \citet{morwood98} \\
P60 & SEDM & 3800-9200  & 100 & \citet{Blagorodnova2018} \\
P200 & DBSP & 3500-10500  & 1000 & \citet{Oke1982} \\
Seimei & KOOLS-IFU & 4100-8900  & 500 & \citet{Matsubayashi2019} \\
VLT & X-Shooter & 3000-24600  & 6000 & \citet{vernet11} \\
\hline
Liverpool Tel. & IO:O & \textit{ugriz} & --- & \citet{Steele2004} \\
P48 & ZTF Camera & \textit{gri} & --- & \citet{Dekany2020} \\
P60 & Rainbow Camera & \textit{gri} & --- & \citet{Blagorodnova2018} \\
Seimei & TriCCS & \textit{gri} & --- & URL\tablefootmark{2} \\
NOT & NOTCam & \textit{JHKs} & --- & URL\tablefootmark{3} \\
NTT & SOFI & \textit{JHKs} & --- & \citet{morwood98} \\
\hline
\end{tabular}
\tablefoot{
\\
\tablefoottext{1}{\url{http://www.not.iac.es/instruments/alfosc/}} \\
\tablefoottext{2}{\url{http://www.o.kwasan.kyoto-u.ac.jp/inst/triccs/index.html}} \\
\tablefoottext{3}{\url{http://www.not.iac.es/instruments/notcam/}} \\
}
\end{table*}

\begin{figure}[h!]
   \centering
   \includegraphics[width=0.5\textwidth]{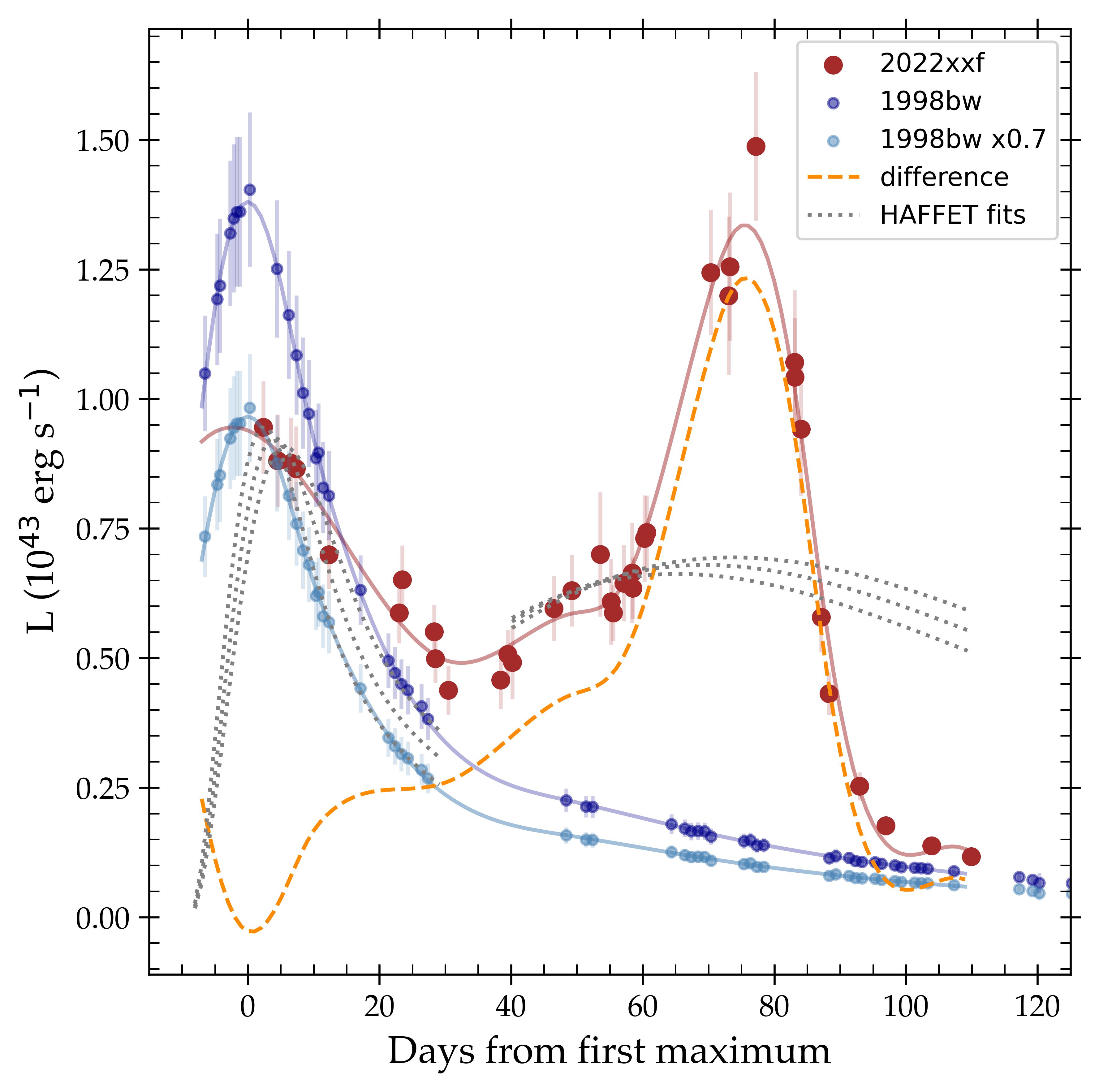} 
   \caption{Bolometric light curves of SN~2022xxf (red points) and SN~1998bw \citep{patat01,clocchiatti11} (blue points; light blue for the scaled-down LC). Solid lines are spline representations of the LCs. The difference between the LC of SN~2022xxf and the scaled-down LC of SN~1998bw is plotted with dashed orange line. Dotted grey lines indicate $^{56}$Ni fits using HAFFET.}
   \label{fig:lcbol}
\end{figure}

\begin{figure*}
\centering
\begin{subfigure}{.5\textwidth}
  \centering
  \includegraphics[width=\linewidth]{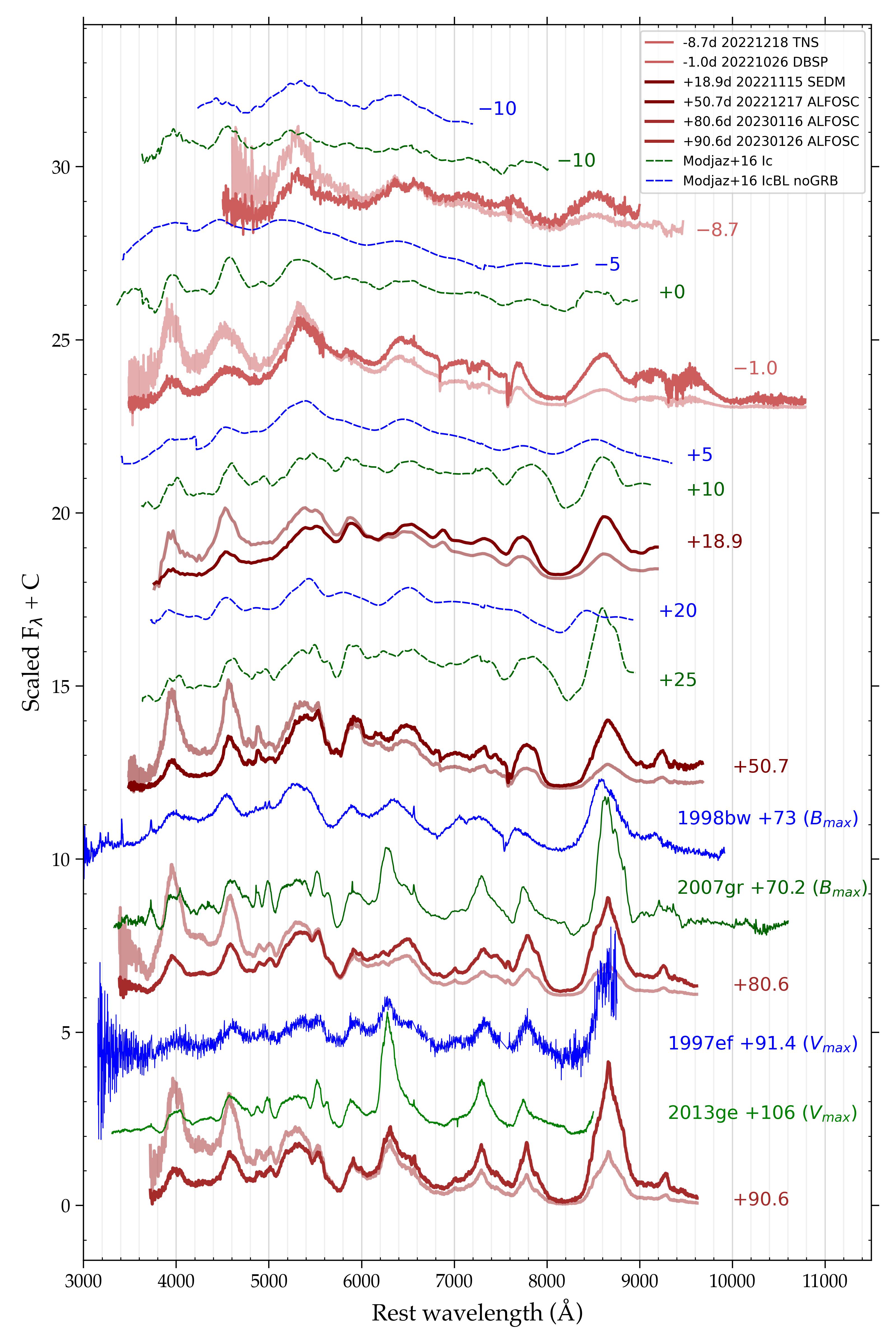}
\end{subfigure}%
\begin{subfigure}{.5\textwidth}
  \centering
  \includegraphics[width=\linewidth]{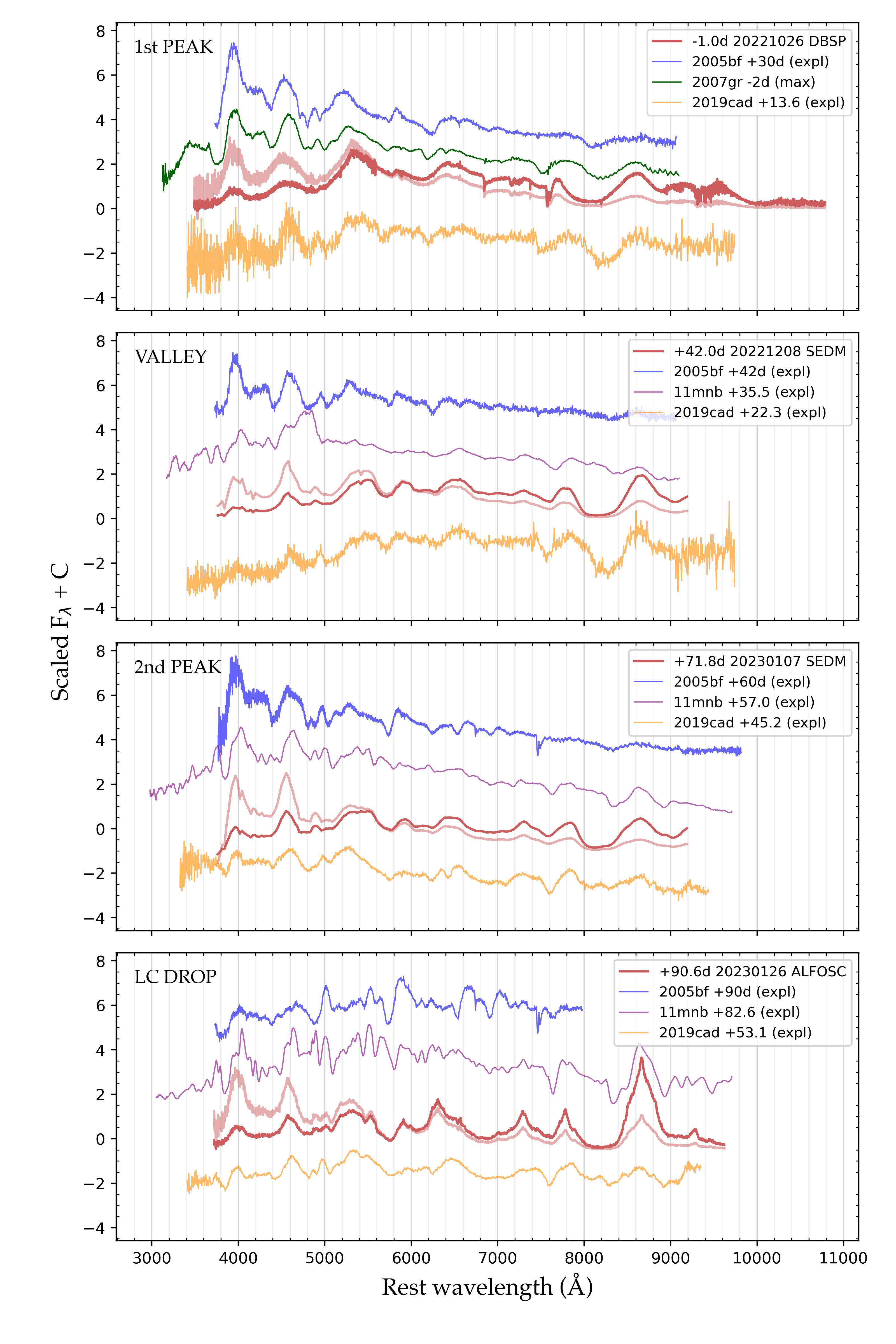}
\end{subfigure}
\caption{(\textit{Left panel}) Spectral comparison of SN~2022xxf to SN Ic (green dashed lines) and GRB-less SN IcBL (blue dashed lines) templates \citep{modjaz16}, and other SNe after +50 days not covered by the templates. The template spectra are flattened, and thus do not represent the correct SED shape -- comparison is intended for identifying similar spectral features. Well-observed Type Ic SNe 2007gr \citep[][{spectrum from \citealt{shivvers19}}]{hunter09} and 2013ge \citep{drout16} are plotted in green solid lines, and SNe IcBL 1998bw \citep{patat01} and 1997ef \citep{modjaz14} {are plotted} in blue solid lines. Spectra of SN~2022xxf are plotted in reddish colors, with the dereddened ones in lighter shades. Phases are in days {relative to the first peak of the light curve}.
(\textit{Right panels}) Spectral comparison of SN~2022xxf with other two-humped objects, during specific LC phases: around the first peak, the `valley' between the peaks, around the second peak, and the fall after the second peak. The LC-peak spectrum of SN~2007gr is shown in the first panel for comparison. 
\label{fig:speccomp}}
\end{figure*}

\begin{figure}[h]
   \centering
   \includegraphics[width=0.5\textwidth]{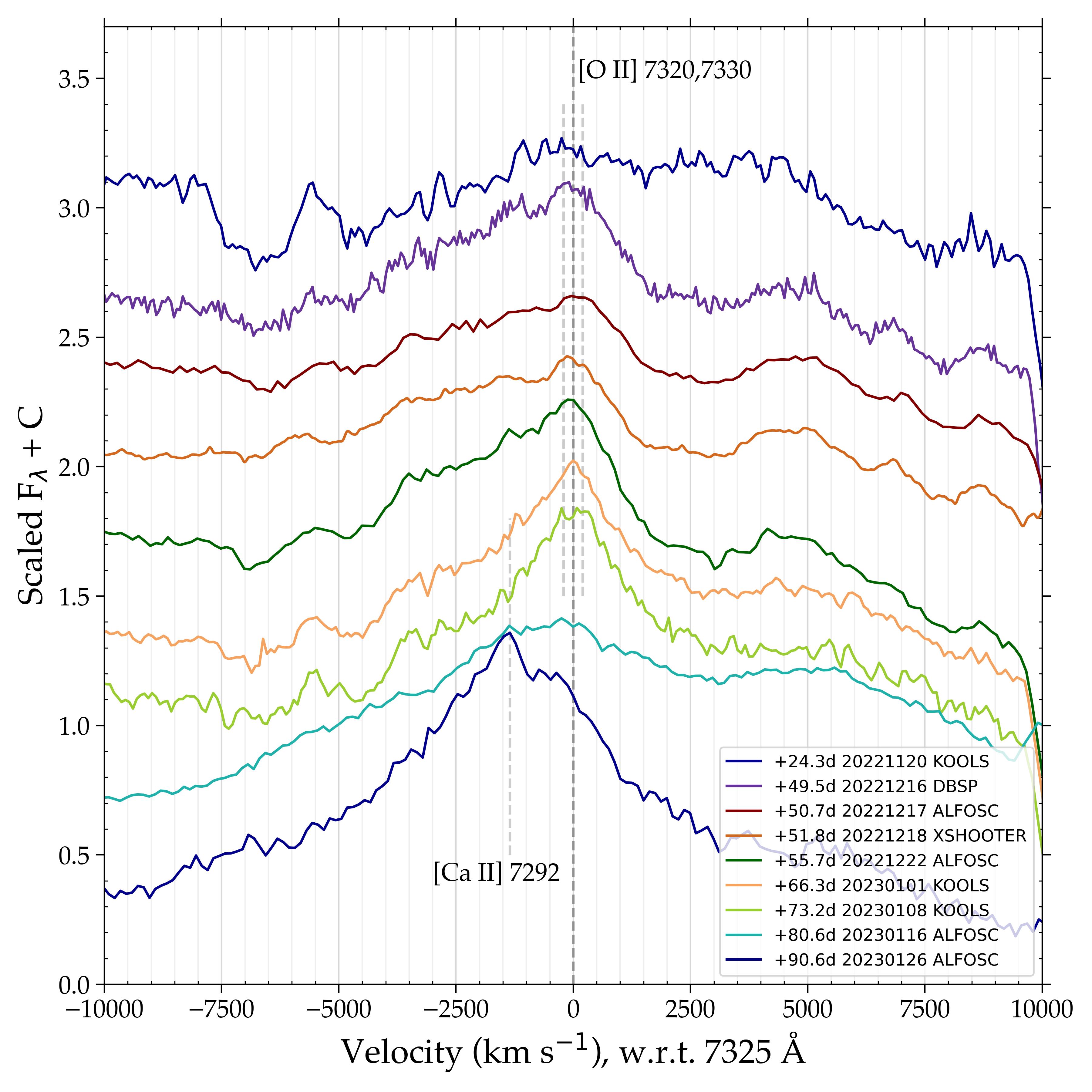} 
   \caption{Evolution of the narrow line at 7325 \AA~ in velocity space. The rest wavelengths of [Ca II] and [O II] are indicated with dashed vertical lines ($\lambda$7324 is almost coincident with zero velocity).}
   \label{fig:line7325}
\end{figure}

\begin{figure*}
\centering
\begin{subfigure}{.33\textwidth}
  \centering
  \includegraphics[width=\linewidth]{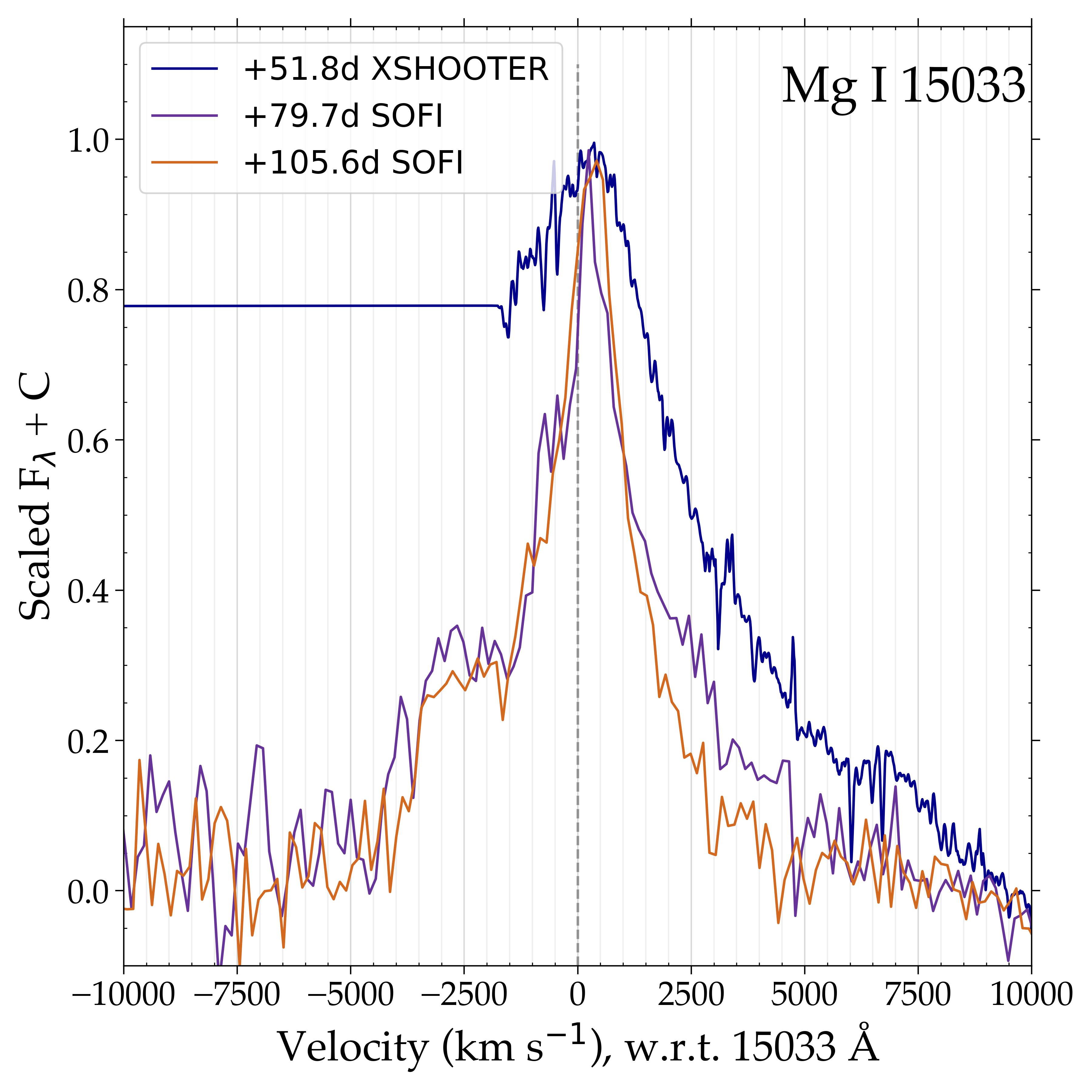}
\end{subfigure}%
\begin{subfigure}{.33\textwidth}
  \centering
  \includegraphics[width=\linewidth]{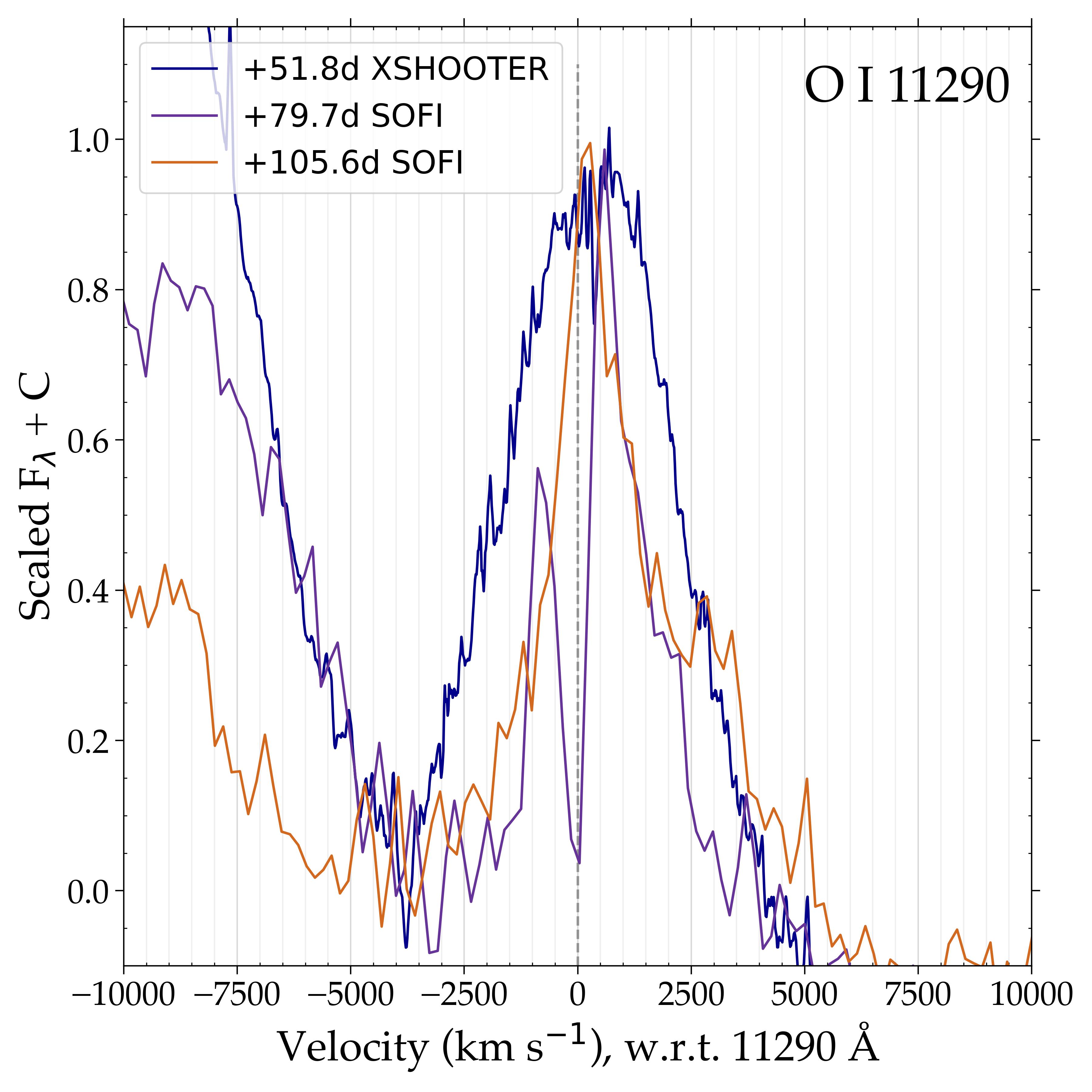}
\end{subfigure}
\begin{subfigure}{.33\textwidth}
  \centering
  \includegraphics[width=\linewidth]{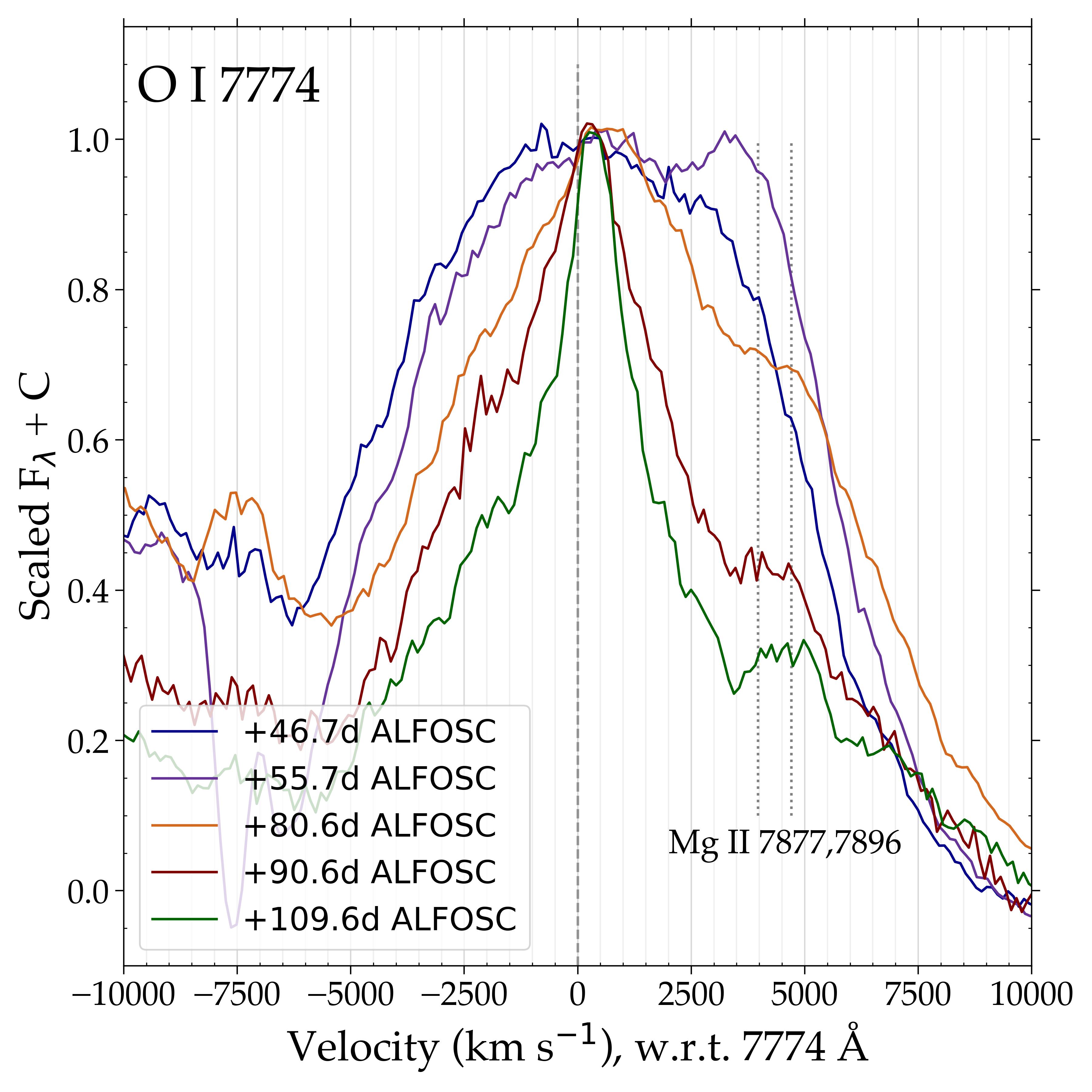}
\end{subfigure}
\caption{Line profile{s} of selected O and Mg emission lines in the optical and NIR.}
\label{fig:lineprofs}
\end{figure*}

\begin{figure*}
   \centering
   \includegraphics[width=\textwidth]{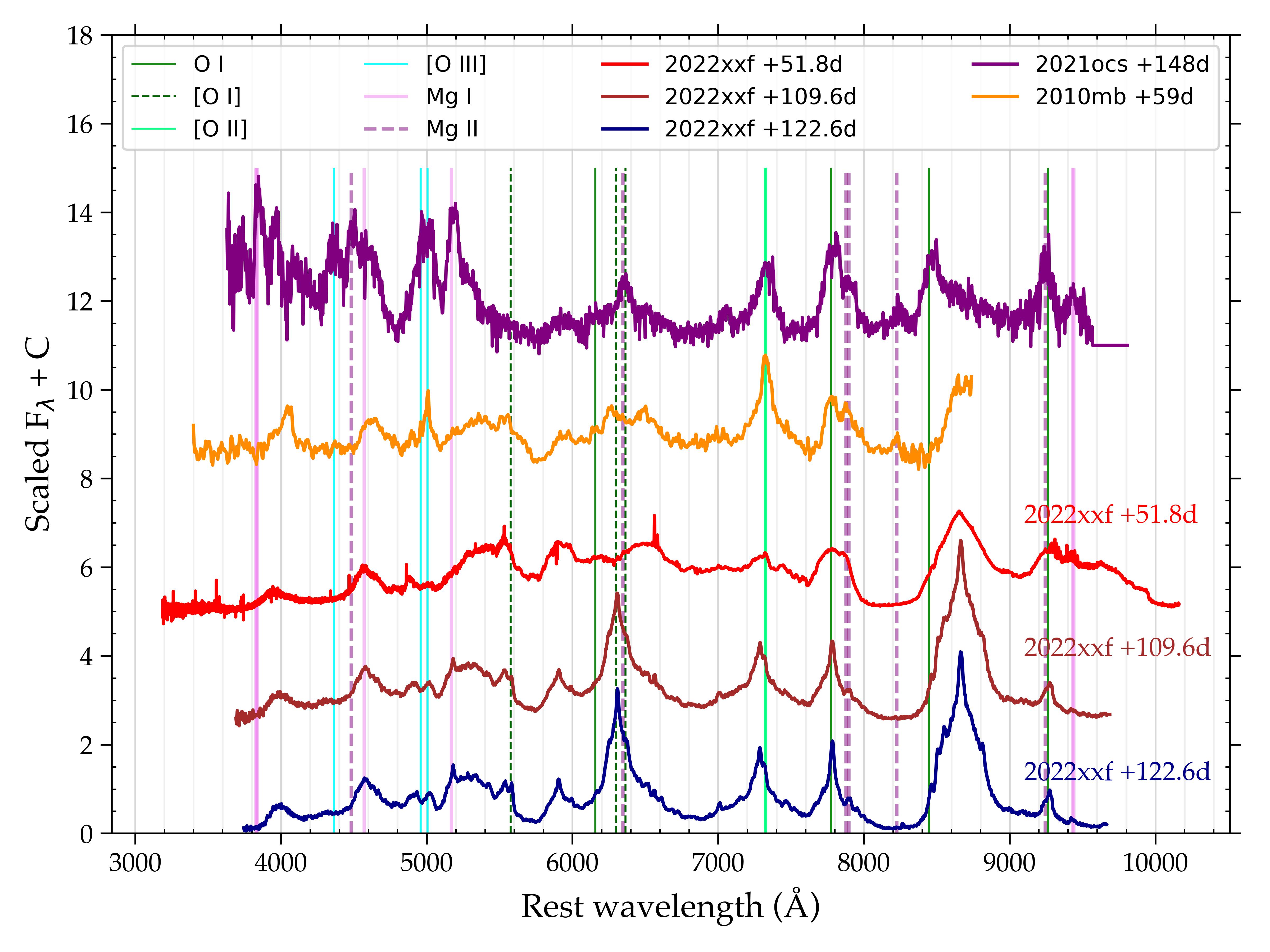} 
   \caption{Spectral comparison of SN~2022xxf to other interacting Type Ic SNe.}
   \label{fig:compneb}
\end{figure*}

\end{appendix}

\end{document}